\documentclass[prd,notitlepage,showpacs,nofootinbib,superscriptaddress]{revtex4-2}

\usepackage[utf8]{inputenc} 

\usepackage{amsmath}
\usepackage{amssymb}
\usepackage{float}
\usepackage{comment}
\usepackage{slashed}
\usepackage[normalem]{ulem}
\usepackage{bbold}
\usepackage{cancel}
\usepackage{graphicx}
\usepackage{booktabs}
\usepackage{multirow}
\usepackage{rotating}
\usepackage{caption}
\usepackage{subcaption}
\usepackage{tabulary}

\newcolumntype{K}[1]{>{\centering\arraybackslash}m{#1}}
\def\gsim{\raise0.3ex\hbox{$\;>$\kern-0.75em\raise-1.1ex\hbox{$\sim\;$}}}
\def\lsim{\raise0.3ex\hbox{$\;<$\kern-0.75em\raise-1.1ex\hbox{$\sim\;$}}}

\DeclareMathOperator{\diag}{diag}

\usepackage[usenames,dvipsnames]{color}

\newcommand {\ignore}[1]{}

\usepackage[colorinlistoftodos]{todonotes}
\usepackage[colorlinks=true,linkcolor=darkblue,citecolor=darkblue,urlcolor=darkblue, pdfborder={0 0 0}]{hyperref}
\usepackage[normalem]{ulem}

\usepackage{caption}
\usepackage{subcaption}
\definecolor{linkcolor}{rgb}{0,0,0.8}

\definecolor{darkgreen}{rgb}{0,0.5,0}
\definecolor{darkred}{rgb}{0.6,0,0}
\definecolor{brown}{rgb}{0.59, 0.29, 0.0}
\definecolor{mightnightblue}{RGB}{25,25,112}
\definecolor{darkblue}{rgb}{0,0,0.8}

\newcommand{\U}{\mathbf{U}}

\newcommand{\Y}{\mathbf{Y}}

\def\Y{\mathbf{Y}}

\def\U{\mathbf{U}}

\newcommand{\AddrCFTP}{%
Departamento de F\'{\i}sica and CFTP, Instituto Superior T\'ecnico, Universidade de Lisboa, Av. Rovisco Pais 1, 1049-001 Lisboa, Portugal}

\usepackage[force]{feynmp-auto}

\usepackage{colortbl}

\begin{document}

\title{Flavored Peccei-Quinn symmetries \\
in the minimal $\nu$DFSZ model}

\author{\textbf{J. R. Rocha}}\email{jose.r.rocha@tecnico.ulisboa.pt}
\affiliation{\AddrCFTP}

\author{\textbf{H.~B. C\^amara}}\email{henrique.b.camara@tecnico.ulisboa.pt}
\affiliation{\AddrCFTP}

\author{\textbf{F.~R. Joaquim}}\email{filipe.joaquim@tecnico.ulisboa.pt }
\affiliation{\AddrCFTP}

\begin{abstract}
\vspace{0.2cm}
\begin{center}
{ \center  \bf ABSTRACT}\\    
\end{center}
We consider a Dine-Fischler-Srednicki-Zhitnitsky~(DFSZ) axion model extended with two right-handed neutrino fields to realize the minimal type-I seesaw. In this $\nu$DFSZ scheme we systematically determine the simplest quark and lepton flavor patterns compatible with masses, mixing and charge-parity violation data, realized by flavored U(1) Peccei-Quinn~(PQ) symmetries. We discuss axion dark matter production in pre and post-inflationary cosmology in this context, and predictions for the axion couplings to photons and fermions. In particular, helioscopes and haloscopes are able to probe our models via their distinct axion-to-photon couplings, while in the quark sector the most stringent constraints on axion-fermion couplings are set by $K^+ \rightarrow \pi^+ + a$. Flavor-violating constraints in the lepton sector are not as relevant as those stemming from Red Giants and Star Cooling SN1987a that restrict the diagonal $ee$ and $\mu\mu$ axion couplings to charged leptons, respectively. We also obtain axion mass bounds for the most interesting models and discuss how minimal flavored PQ symmetries provide a natural framework to suppress flavor-violating couplings.
\end{abstract}

\maketitle
\noindent

\section{Introduction}
\label{sec:intro}

The discovery of the Higgs boson by the ATLAS and CMS collaborations~\cite{ATLAS:2015yey} at the CERN Large Hadron Collider~(LHC) shed light to the mechanism responsible for providing mass to the $W$ and $Z$ bosons mediating weak interactions to quarks and charged leptons in the Standard Model (SM). While the SM weak sector is remarkably predictive and thoroughly tested in experiments, the flavor sector remains puzzling, offering no fundamental explanation for the observed pattern of fermion masses and mixing. For quarks, the mismatch between weak and mass eigenstates is encoded in the Cabibbo-Kobayashi-Maskawa~(CKM) matrix~\cite{Cabibbo:1963yz,Kobayashi:1973fv} which violates the charge-parity~(CP) symmetry through a complex phase. In contrast, lepton mixing is absent since neutrinos are massless in the SM. The groundbreaking discovery of neutrino oscillations~\cite{McDonald:2016ixn,Kajita:2016cak} not only established the fact that neutrinos are massive particles but also revealed that lepton mixing is non trivial. These observations comprise a clear evidence for physics beyond the SM~(BSM), paving the way for new phenomenological studies at the intensity, energy and cosmic frontiers of particle physics.

Unlike the other fermions, neutrinos can be their own antiparticles~\cite{Majorana:1937vz}, a possibility that has motivated numerous theoretical models for Majorana neutrino masses in the context of ultraviolet completions of the SM that realize the effective Weinberg operator~\cite{Weinberg:1980wa}. Among these, the seesaw mechanism~\cite{Minkowski:1977sc,Gell-Mann:1979vob,Yanagida:1979as,Schechter:1980gr,Glashow:1979nm,Mohapatra:1979ia} is the most widely studied framework for explaining the suppression of neutrino masses. The minimal type-I seesaw is particularly appealing, as it requires only two right-handed (RH) neutrinos to accommodate neutrino oscillation data~\cite{Frampton:2002qc,Ibarra:2003up,Harigaya:2012bw,Rink:2016vvl,Shimizu:2017fgu,Barreiros:2018ndn,Barreiros:2018bju,Barreiros:2020mnr}. Within the three-neutrino mixing paradigm, global fits provide current up-to-date values of the neutrino observables, some of which known to an accuracy below the percent level~\cite{deSalas:2020pgw,Esteban:2024eli,Capozzi:2025wyn}. Nonetheless, these experiments are not sensitive to the absolute neutrino mass scale leaving open the possibility for a massless light neutrino, a feature of the minimal type-I seesaw. Furthermore, it is still unknown whether neutrinos are Majorana or Dirac, what their mass ordering is and whether or not there is leptonic CP violation.

The flavor puzzle has motivated a plethora of SM extensions featuring additional fields and new symmetries~\cite{Ishimori:2010au,Morisi:2012fg,King:2017guk,Petcov:2017ggy,Feruglio:2019ybq,Xing:2020ijf}. In this work, we focus on global U(1) Abelian symmetries that give rise to texture zeros in the Yukawa~\cite{Grimus:2004hf,Dighe:2009xj,Adhikary:2009kz,Dev:2011jc,GonzalezFelipe:2014zjk,Samanta:2015oqa,Kobayashi:2018zpq,Rahat:2018sgs,Nath:2018xih,Barreiros:2018ndn,Barreiros:2018bju,Correia:2019vbn,Camara:2020efq,Barreiros:2020gxu,Barreiros:2022aqu,Rocha:2024twm} and mass matrices~\cite{Ludl:2014axa,Ludl:2015lta,GonzalezFelipe:2014zjk,Cebola:2015dwa}. When implemented within the SM, these symmetries generate fermion mass and mixing structures that are ruled out by current experimental data~\cite{Correia:2019vbn,Camara:2020efq,Rocha:2024twm}. The simplest extension of the SM that successfully accommodates U(1) flavor symmetries is the two-Higgs-doublet model (2HDM)~\cite{Branco:2011iw}. Namely, they have been employed in 2HDM frameworks to address the flavor puzzles in the quark~\cite{Ferreira:2010ir} and lepton~\cite{GonzalezFelipe:2014zjk,Correia:2019vbn,Camara:2020efq} sectors separately. More recently, a combined analysis of both sectors was performed~\cite{Rocha:2024twm}. It was shown that Abelian flavor symmetries provide a natural mechanism to suppress and control flavor-changing neutral couplings (FCNCs), which are tightly constrained by quark flavor observables and also induce contributions to charged-lepton flavor-violating processes, currently being targeted in ongoing experimental searches.

Another longstanding aesthetic issue of the SM concerns the so-called strong CP phase $\bar{\theta}$, which encodes CP violation in Quantum Chromodynamics~(QCD). Experimental searches for the neutron electric dipole moment place a stringent upper bound on this parameter, namely $|\bar{\theta}| < 10^{-10}$~\cite{PhysRevD.92.092003,PhysRevLett.97.131801}. Although this suggests that QCD preserves CP symmetry to a high degree, the SM does not offer a theoretical justification for the smallness of $\bar{\theta}$, a puzzle known as the strong CP problem. One of the most popular and elegant solutions to this issue is the Peccei–Quinn (PQ) mechanism~\cite{PhysRevLett.38.1440,PhysRevD.16.1791}, which introduces a global, QCD-anomalous $\text{U}(1)_{\text{PQ}}$ symmetry. Its spontaneous breaking gives rise to a pseudo-Goldstone boson~(GB) — the axion $a$~\cite{PhysRevLett.40.223,PhysRevLett.40.279} — which acquires mass through non-perturbative QCD effects. The ground state of the axion potential is such that it dynamically drives $\bar{\theta} \to 0$, thereby solving the strong CP problem. Among the SM extensions where the PQ mechanism can be implemented, two main frameworks stand out: Dine–Fischler–Srednicki–Zhitnitsky (DFSZ)~\cite{Zhitnitsky:1980tq,Dine:1981rt} and Kim–Shifman–Vainshtein–Zakharov (KSVZ)~\cite{Kim:1979if,Shifman:1979if} invisible axion models. In the DFSZ scenario, SM quarks are charged under $\text{U}(1)_{\text{PQ}}$, whereas in the KSVZ model, the symmetry is assigned to additional heavy exotic quarks (see Ref.~\cite{DiLuzio:2020wdo} for a review). The original DFSZ model features an extended scalar sector comprising two Higgs doublets and a complex scalar singlet. The axion is predominantly the phase of the singlet, whose vacuum expectation value (VEV) sets the PQ-breaking scale. A strong hierarchy between the singlet and doublet VEVs is required to render the axion effectively “invisible”. In contrast, the earlier PQ–Weinberg–Wilczek (PQWW) model~\cite{PhysRevLett.38.1440,PhysRevD.16.1791,PhysRevLett.40.223,PhysRevLett.40.279}, which included only two Higgs doublets, has been excluded by experimental bounds from rare meson decays such as $K^+ \rightarrow \pi^+ + \text{invisible}$~\cite{Wilczek:1977zn,Donnelly:1978ty,Hall:1981bc,Frere:1981cc,Asano:1981nh,Zehnder:1981qn,Bardeen:1986yb}.

Axion frameworks provide a gateway to address multiple BSM phenomena in a complementary way. Notably, axion relics can be generated non-thermally in the early Universe via the so-called misalignment mechanism, thus accounting for the observed dark matter (DM) relic density~\cite{Preskill:1982cy,Abbott:1982af,Dine:1982ah}. Furthermore, the DFSZ framework has been studied in connection with the flavor puzzle in the quark sector, with the PQ symmetry promoted to a flavor symmetry in the same spirit as Abelian flavor symmetries~\cite{Davidson:1981zd,Celis:2014zaa,Celis:2014jua,Celis:2014iua,Cox:2023squ,delaVega:2021ugs}. In Ref.~\cite{Cox:2023squ}, the authors provide a classification of flavored DFSZ models that are free of domain walls (DWs), i.e. featuring a DW number $N_{\text{DW}} = 1$, focusing exclusively on the quark sector. It would be interesting to systematically study axion flavor models that also incorporate neutrino mass generation. Namely, the axion–neutrino connection through the type-I seesaw mechanism, where the VEV of the scalar singlet provides the Majorana masses for RH neutrinos, has been explored in the context of the DFSZ axion~\cite{Volkas:1988cm,Clarke:2015bea,Sopov:2022bog,RVolkas:2023jiv,Matlis:2023eli}, giving rise to the so-called $\nu$DFSZ model. Similar constructions have also been considered within the KSVZ scenario~\cite{Salvio:2015cja,Ballesteros:2016euj,Ballesteros:2016xej}. More recently, within the KSVZ framework, exotic colored mediators were used to generate radiatively both Majorana~\cite{Batra:2023erw} and Dirac~\cite{Batra:2025gzy} neutrino masses.

In this work, we extend the analysis of our previous paper~\cite{Rocha:2024twm}, by considering flavored PQ symmetries in the minimal $\nu$DFSZ model which features two-Higgs doublets and a complex singlet in the scalar sector and two RH neutrinos generating type-I seesaw neutrino masses. The PQ symmetry necessary to implement the axion solution to the strong CP problem will act as our flavor symmetry. Namely, we are interested in scenarios where the PQ symmetry maximally restricts the quark and lepton mass matrices, such that these contain the least possible number of independent parameters required to reproduce the fermion masses, mixing and CP violation data, while satisfying all relevant phenomenological constraints. In contrast to Ref.~\cite{Cox:2023squ}, here we systematically classify for the first time the minimal complete quark and lepton flavor scenarios considering cases that may feature $N_{\text{DW}} \neq 1$ and incorporate neutrino masses via the seesaw mechanism. The lepton sector plays a crucial role in axion coupling phenomenology. The paper is organized as follows. In Sec.~\ref{sec:model}, we present the $\nu$DFSZ model which supports the systematic analysis of Sec.~\ref{sec:symmetries}, where we identify the most restrictive quark and lepton flavor structures which are compatible with data and realizable by flavored PQ symmetries. Next, in Sec.~\ref{sec:axionpheno} we study the axion phenomenology of our models, starting with a discussion on axion DM in pre and post-inflationary cosmology in Sec.~\ref{sec:axiondarkmatter}. We show how our models can be experimentally distinguished through their distinct predictions for the axion-to-photon coupling and axion-to-fermion flavor-violating couplings in Secs.~\ref{sec:axionphoton} and \ref{sec:axionflavorviolating}, respectively. Finally, we draw our conclusions in Sec.~\ref{sec:concl}.

\section{The flavored $\nu$DFSZ model}
\label{sec:model}

In our minimal $\nu$DFSZ, the scalar field content consists of two-Higgs doublets, $\Phi_{1,2} \sim (1,2,1)$, and a complex singlet scalar, $\sigma \sim (1,1,0)$, with transformation properties defined under the SM gauge group ${\rm G}_{\rm SM}=\text{SU}(3)_c \times \text{SU}(2)_L \times \text{U}(1)_Y$. The SM fermion field content is extended with two RH neutrinos, $\nu_{R_{1,2}}\sim(1,1,0)$. A global U(1)$_{\text{PQ}}$ is imposed, under which the various fields generically transform as:
\begin{equation}
\begin{aligned}  
&\Phi_k \rightarrow \exp\left({i \zeta \chi_{k} }\right) \Phi_k \;,& &q_{L_\alpha} \rightarrow  \exp\left({i \zeta \chi_{q_\alpha}^L }\right) q_{L \alpha} \;,&  &d_{R_\alpha} \rightarrow \exp\left({i \zeta \chi_{d_\alpha}^R }\right) d_{R \alpha} 
\;,&  & u_{R \alpha} \rightarrow \exp\left({i \zeta \chi_{u_\alpha}^R }\right) u_{R \alpha}
\; ,&  
\\
 &\sigma \rightarrow \exp\left({i \zeta \chi_\sigma }\right) \sigma \;,& &\ell_{L \alpha} \rightarrow  \exp\left({i \zeta \chi_{\ell_\alpha}^L }\right) \ell_{L \alpha} \;,& &e_{R \alpha} \rightarrow \exp\left({i \zeta \chi_{e_\alpha}^R }\right) e_{R \alpha} 
 \;,&  & \nu_{R j} \rightarrow  \exp\left({i \zeta \chi_{\nu_j}^R }\right) \nu_{R j}
 \; ,& 
\label{eq:PQSym}
\end{aligned}   
\end{equation}
where $k = 1,2 $ labels the Higgs doublets, $\alpha = 1,2,3$ the SM fermion generations and $j =1,2$ the number of RH neutrinos. The PQ charges are denoted by $\chi$'s and $\zeta$ is a continuous parameter. As usual, the left-handed~(LH) quark doublets, RH down-quark singlets, and RH up-quark singlets are defined as $q_L = \left(u_L, d_L\right)^T$, $d_R$, and $u_R$, respectively. Similarly, in the leptonic sector, the LH lepton doublets and RH charged-lepton singlets are $\ell_L = \left(\nu_L, e_L\right)^T$ and $e_R$, respectively.

Given the field transformation properties under ${\rm G}_{\rm SM}$ and the PQ symmetries of Eq.~\eqref{eq:PQSym}, the scalar potential is:
\begin{align}
    V(\Phi_1,\Phi_2,\sigma) &= \mu_{11}^2 \Phi_1^\dagger \Phi_1 + \mu_{22}^2 \Phi_2^\dagger \Phi_2 + \frac{\lambda_1}{2} \left(\Phi_1^\dagger \Phi_1\right)^2 + \frac{\lambda_2}{2} \left(\Phi_2^\dagger \Phi_2\right)^2 + \lambda_3 \left(\Phi_1^\dagger \Phi_1\right)\left(\Phi_2^\dagger \Phi_2\right) + \lambda_4 \left(\Phi_1^\dagger \Phi_2\right)\left(\Phi_2^\dagger \Phi_1\right) \nonumber
    \\
    & + \mu_{\sigma}^2 |\sigma|^2 + \frac{\lambda_\sigma}{2} |\sigma|^4 + \lambda_{1 \sigma} \Phi_1^\dagger \Phi_1 |\sigma|^2
    + \lambda_{2 \sigma} \Phi_2^\dagger \Phi_2 |\sigma|^2
    +
    \left[V_{\text{PQ}}(\Phi_1,\Phi_2,\sigma) +\text{H.c.} \right] 
    \; ,
    \label{eq:VpotentialDFSZ}
\end{align}
where, depending on the scalar field PQ charges, two classes of models can be identified for which $V_{\text{PQ}}$ is given by:
\begin{align}
 {\rm Cubic \,Model:}\,&\;  V_{\text{PQ}}(\Phi_1,\Phi_2,\sigma) =
    \kappa \Phi_2^\dagger \Phi_1 \sigma \; \;,\;\; 
    -\chi_{2}+\chi_{1} + \chi_{\sigma} = 0 \  \text{(mod $2\pi$)}
    \; , \nonumber \\
    {\rm Quartic \,Model:}\,& \;
   V_{\text{PQ}}(\Phi_1,\Phi_2,\sigma) = \lambda \Phi_2^\dagger \Phi_1 \sigma^2
    \; \;,\;\;
    \; \; 
   -\chi_{2}+\chi_{1} + 2\chi_{\sigma} = 0 \  \text{(mod $2\pi$)} \; .
\label{eq:VPQoperator}
\end{align}
Notice that $V_{\text{PQ}}$ is a necessary term in the scalar sector since, if neither of the PQ charge relations shown above are satisfied, the full scalar potential exhibits a U(1)$^3$ symmetry. This results in an additional massless neutral GB, besides the one corresponding to the longitudinal component of the $Z$-boson, and the axion, which gets a mass via non-perturbative QCD effects, making it a pseudo-GB. 

We parametrize the scalar fields as,
\begin{equation}
     \Phi_{1,2} = \frac{1}{\sqrt{2} }
             \begin{pmatrix} \phi_{1,2}^+ \\ \left(v_{1,2} + \rho_{1,2}\right) e^{i a_{1,2}/v_{1,2}} \end{pmatrix}
    \; , \quad 
     \sigma = \frac{1}{\sqrt{2} } \left(v_\sigma + \rho_\sigma \right) e^{i a_\sigma/v_\sigma}  \; ,
\label{eq:parametrisation}
\end{equation} 
where the charged components of the Higgs doublets are $\phi_{1,2}^+$, while $a_{1,2,\sigma}$ denote the neutral-scalar phases, and $\rho_{1,2,\sigma}$ the radial modes. The VEVs of the $\Phi_{1,2}$ ($\sigma$) neutral component are $v_{1,2}$ ($v_\sigma$). With this setup, the physical axion is given by,
\begin{equation}
    a = \frac{1}{v_a} \sum_{i=1,2,\sigma} \chi_i a_i v_i \; , \quad 
    v_a^2 = \sum_{i=1,2,\sigma} \chi_i^2 v_i^2 \; .
    \label{eq:axionandvev}
\end{equation}
Note that, assuming a strong hierarchy among the singlet and doublet VEVs $v_\sigma \gg v_1,v_2$, the physical axion and $v_a$ essentially stem from $\sigma$ and $v_\sigma$, respectively. Imposing that the PQ current is orthogonal to the hypercharge current enforces $ v_1^2\chi_{1} + v_2^2\chi_{2}=0$ and guarantees absence of kinetic mixing between the axion and the $Z$-boson. To satisfy this orthogonality condition, along with PQ invariance of $V$, we set:
\begin{align}
    \chi_{1} = -\sin^2 \beta \equiv - s_\beta^2 \; , \quad 
    \chi_{2} = \cos^2 \beta \equiv c_\beta^2 \; , \quad 
     \chi_{\sigma} = 
    \begin{cases}
        \;\; 1 \quad &\text{(Cubic Model)} \\
        \;\; \dfrac{1}{2} \quad &\text{(Quartic Model)} 
    \end{cases} \; , \; \tan \beta = v_2/v_1\,,
    \label{eq:PQsymmetry}
\end{align}
without loss of generality.

The Yukawa and mass terms invariant under ${\rm G}_{\rm SM}\times {\rm U}(1)_{\rm PQ}$ are:
\begin{align}
    -\mathcal{L}_\mathrm{Yuk.} & = \overline{q_L}\left(\Y^d_1 \Phi_1 + \Y^d_2 \Phi_2\right)d_R 
    + \overline{q_L} \left(\Y^u_1 \Tilde{\Phi}_1 + \Y^u_2 \Tilde{\Phi}_2\right) u_R + \overline{\ell_L}\left(\Y^e_1 \Phi_1 + \Y^e_2 \Phi_2\right)e_R \, \nonumber
    \\
    &+ \overline{\ell_L}\left(\Y^{D \ast}_{1} \Tilde{\Phi}_1 + \Y^{D \ast}_{2} \Tilde{\Phi}_2\right)\nu_R 
    + \frac{1}{2} \overline{\nu_R} \left(
    \mathbf{M}^0_R +  \Y^R_1 \sigma + \Y^R_2 \sigma^\ast
    \right) \nu_R^c
    + \mathrm{H.c.} \; ,
    \label{eq:Lyuk2hdm}
\end{align}
where $\tilde{\Phi}_{1,2} = i \tau_2 \Phi_{1,2}^*$, with $\tau_2$ being the complex Pauli matrix, and $\nu_R^c \equiv C \overline{\nu_R}^T$ ($C$ is the charge conjugation matrix). The Yukawa matrices $\mathbf{Y}^{f (D)}_{1,2}$ ($f = d, u, e$) are $3 \times 3$ (or $3 \times 2$) general complex matrices, while $\mathbf{Y}^R_{1,2}$ and the bare Majorana mass matrix $\mathbf{M}_R^0$ are $2 \times 2$ symmetric matrices. Invariance under the PQ symmetry requires,
\begin{equation}
\begin{aligned}    
    (\mathbf{Y}_{1,2}^{f})_{\alpha \beta} &= \exp\left[{i (\Theta_{1,2}^{f})_{\alpha \beta}} \right](\mathbf{Y}_{1,2}^{f})_{\alpha \beta} \; , \quad &(\mathbf{Y}_{1,2}^{D})_{\alpha j} &= \exp\left[{i (\Theta_{1,2}^{D})_{\alpha j}} \right](\mathbf{Y}_{1,2}^{D})_{\alpha j} \; ,
    \\
    (\mathbf{Y}_{1,2}^{R})_{i j} &= \exp\left[{i (\Theta_{1,2}^{R})_{i j}} \right](\mathbf{Y}_{1,2}^{R})_{i j} \; , \quad &(\mathbf{M}_{R}^0)_{ij} &= \exp\left[{i (\Theta_0^R)_{ij}}\right] (\mathbf{M}^{0}_R)_{ij} \; , 
\label{eq:Invariance}
\end{aligned}    
\end{equation}
where $\alpha,\beta$ run over the three SM fermion generations, $i,j=1,2$, and
\begin{gather}
(\Theta^{d}_{1,2})_{\alpha\beta}  =\chi_{d_\beta}^R-\chi_{q_\alpha}^L+\chi_{1,2} \; , \quad 
(\Theta^u_{1,2})_{\alpha\beta}=\chi_{u_\beta}^R-\chi_{q_\alpha}^L-\chi_{ 1,2} \; , \nonumber
\\
(\Theta^e_{1,2})_{\alpha\beta}=\chi_{e_\beta}^R-\chi_{q_\alpha}^L+\chi_{ 1,2} \; , \quad
(\Theta^D_{1,2})_{\alpha j} =\chi_{\nu_j}^R-\chi_{\ell_\alpha}^L - \chi_{ 1,2} \; , \nonumber
\\
(\Theta^R_0)_{ij}=\chi_{\nu_j}^R +\chi_{\nu_i}^R  \; , \quad 
(\Theta^R_1)_{ij}=\chi_{\nu_j}^R +\chi_{\nu_i}^R - \chi_{\sigma} \; , \quad   
(\Theta^R_2)_{ij}=\chi_{\nu_j}^R +\chi_{\nu_i}^R + \chi_{\sigma} \;  .
\label{eq:canonicalcharges}
\end{gather}
Alternatively, Eqs.~\eqref{eq:Invariance} can be reformulated as follows:
\begin{equation}
\begin{aligned}
(\Theta^{f}_{1,2})_{\alpha \beta} & \neq 0 \  \text{(mod $2\pi$)}  \Leftrightarrow  (\mathbf{Y}^{f}_{1,2})_{\alpha \beta} =  0 \; , \quad  &(\Theta^{D}_{1,2})_{\alpha j} &\neq 0 \  \text{(mod $2\pi$)} \Leftrightarrow (\mathbf{Y}^{D}_{1,2})_{\alpha j} =  0 \,,
\\
(\Theta^{R}_{1,2})_{i j} & \neq 0 \  \text{(mod $2\pi$)}  \Leftrightarrow  (\mathbf{Y}^{R}_{1,2})_{i j} =  0 \; , \quad &(\Theta^R_{0})_{ij} &\neq 0 \  \text{(mod $2\pi$)} \Leftrightarrow (\mathbf{M}_R^0)_{ij} =  0 \;.
\label{eq:YukawaZeros}
\end{aligned}
\end{equation}
Thus, the PQ symmetry may forbid certain Yukawa and bare mass terms in Eq.~\eqref{eq:Lyuk2hdm}, leading to texture-zero patterns for $\mathbf{Y}^{f,D,R}_{1,2}$ and $\mathbf{M}^{0}_R$.

Spontaneous symmetry breaking is triggered when the neutral components of the Higgs doublets and the scalar singlet acquire nonzero VEVs [see Eq.~\eqref{eq:parametrisation}], resulting in the fermion-mass Lagrangian:
\begin{align}
    -\mathcal{L}_{\text {mass }}=\overline{d_L} \mathbf{M}_{d} d_R+\overline{u_L} \mathbf{M}_{u} u_R+\overline{e_L} \mathbf{M}_{e} e_R+\overline{\nu_L} \mathbf{M}_D^* \nu_R+\frac{1}{2} \overline{\nu_R} \mathbf{M}_R \nu_R^c+\text {H.c.}  \; ,
        \label{eq:Mass2hdm}
\end{align}
with
\begin{align}
    \mathbf{M}_{f}=\frac{v_1}{\sqrt{2}} \mathbf{Y}^{f}_1 + \frac{v_2 }{\sqrt{2}} \mathbf{Y}^{f}_2 
    \;,\; 
    \quad \mathbf{M}_D=\frac{v_1}{\sqrt{2}} \mathbf{Y}^{D}_1 + \frac{v_2 }{\sqrt{2}} \mathbf{Y}^{D}_2
    \; , \;
    \quad \mathbf{M}_R=\mathbf{M}_R^0+\frac{v_\sigma}{\sqrt{2}}\left(\mathbf{Y}^R_1+\mathbf{Y}^R_2 \right) \; .
\label{Eq:MassMatrices}
\end{align}
\begin{table}[t!]
\renewcommand*{\arraystretch}{1.5}
\begin{minipage}[b]{.4\textwidth}
\centering
\begin{tabular}{|c c|}
        \hline 
        Parameter & Best fit $\pm 1 \sigma$ \\
        \hline 
        $m_d (\times \; \text{MeV})$ \; \; & $4.67^{+0.48}_{-0.17}$ \\
        $m_s (\times \; \text{MeV})$ \; \;& $93.4^{+8.6}_{-3.4}$  \\
        $m_b (\times \; \text{GeV})$ \; \;& $4.18^{+0.03}_{-0.02}$ \\
        $m_u (\times \; \text{MeV})$ \; \;& $2.16^{+0.49}_{-0.26}$ \\
        $m_c (\times \; \text{GeV})$ \; \;& $1.27 \pm 0.02$  \\
        $m_t (\times \; \text{GeV})$\; \; & $172.69 \pm 0.30$ \\ 
        $\theta_{12}^q (^\circ)$ \; \;& $13.04\pm0.05$ \\
        $\theta_{23}^q (^\circ)$ \; \;& $2.38\pm0.06$ \\
        $\theta_{13}^q (^\circ)$ \; \;& $0.201\pm0.011$ \\
        $\delta^q (^\circ)$ \; \;& $68.75\pm4.5$ \\
        \hline 
\end{tabular}
\end{minipage}
\begin{minipage}[b]{.55\textwidth}
\centering
\begin{tabular}{|c c|}  
        \hline 
        Parameter  & Best Fit $\pm 1 \sigma$ \\ \hline
        $m_e (\times \; \text{keV})$ \; \; & $510.99895000\pm 0.00000015$ \\
        $m_\mu (\times \; \text{MeV})$ \; \;& $105.6583755\pm 0.0000023$   \\
        $m_\tau (\times \; \text{GeV})$ \; \;& $1.77686\pm 0.00012$ \\
        $\Delta m_{21}^2 \left(\times 10^{-5} \ \text{eV}^2\right)$ \; \; & $7.50^{+0.22}_{-0.20}$  \\
        $\left|\Delta m_{31}^2\right| \left(\times 10^{-3}  \ \text{eV}^2\right) [\text{NO}]$ \; \; & $2.55^{+0.02}_{-0.03}$  \\
        $\left|\Delta m_{31}^2\right| \left(\times 10^{-3} \ \text{eV}^2\right) [\text{IO}]$ \; \; & $2.45^{+0.02}_{-0.03}$ \\
        $\theta_{12}^\ell (^\circ)$ \; \; & $34.3\pm1.0$ \\
        $\theta_{23}^\ell (^\circ) [\text{NO}]$ \; \; & $49.26\pm0.79$ \\
        $\theta_{23}^\ell (^\circ) [\text{IO}]$ \; \; & $49.46^{+0.60}_{-0.97}$ \\
        $\theta_{13}^\ell (^\circ) [\text{NO}]$ \; \; & $8.53^{+0.13}_{-0.12}$ \\
        $\theta_{13}^\ell (^\circ) [\text{IO}]$ \; \; & $8.58^{+0.12}_{-0.14}$ \\
        $\delta^\ell  (^\circ) [\text{NO}]$ \; \; & $194^{+24}_{-22}$ \\
        $\delta^\ell  (^\circ) [\text{IO}]$ \; \; & $284^{+26}_{-28}$ \\
        \hline 
\end{tabular}
\end{minipage}
\caption{(Left) Current quark data: masses, mixing angles and Dirac CP phase~\cite{ParticleDataGroup:2022pth}. (Right) Current lepton data: charged-lepton 
masses~\cite{ParticleDataGroup:2022pth}, neutrino mass-squared differences, mixing angles and Dirac CP phase, obtained from the global fit of neutrino oscillation data of Ref.~\cite{deSalas:2020pgw} (see also Refs.~\cite{Esteban:2024eli} and~\cite{Capozzi:2025wyn}).}
\label{tab:data}
\end{table}
The quark mass matrices can be brought to the quark physical basis through the unitary transformations
\begin{equation}
\begin{aligned}
    &d_{L,R} \rightarrow \mathbf{U}_{L,R}^d d_{L,R} &\Rightarrow& &\mathbf{U}_L^{d \dagger} \mathbf{M}_d \mathbf{U}_R^d = \mathbf{D}_d = \text{diag}(m_d,m_s,m_b)& \; , 
    \\
    &u_{L,R} \rightarrow \mathbf{U}_{L,R}^u u_{L,R} &\Rightarrow& &\mathbf{U}_L^{u \dagger} \mathbf{M}_u \mathbf{U}_R^u = \mathbf{D}_u = \text{diag}(m_u,m_c,m_t)& \; ,
    \label{eq:massdiag}
\end{aligned}
\end{equation}
where $m_{d,s,b}$ and $m_{u,c,t}$ denote the physical down- and up-type quark masses. The above unitary matrices are obtained by diagonalizing the Hermitian matrices $\mathbf{H}_{d,u}= \mathbf{M}_{d,u} \mathbf{M}_{d,u}^\dagger$ and $\mathbf{H}_{d,u}^\prime= \mathbf{M}_{d,u}^\dagger \mathbf{M}_{d,u}$ as
\begin{align}
    \mathbf{U}_L^{d,u \dagger} \mathbf{H}_{d,u} \mathbf{U}_L^{d,u} = \mathbf{U}_R^{d,u \dagger} \mathbf{H}_{d,u}^\prime \mathbf{U}_R^{d,u} = \mathbf{D}_{d,u}^2 \; ,
    \label{eq:hermitianmassdiag}
\end{align}
yielding the CKM quark mixing matrix $\mathbf{V}$,
\begin{align}
    \mathbf{V} = \mathbf{U}_L^{u \dagger} \mathbf{U}_L^{d} \;,
    \label{eq:CKM}
\end{align}
for which we adopt the standard parametrization~\cite{ParticleDataGroup:2022pth}
\begin{align}
\mathbf{V}  = \begin{pmatrix}  c_{12}^q c_{13}^q & s_{12}^q c_{13}^q & s_{13}^q e^{-i\delta^q}  \\
- s_{12}^q c_{23}^q - c_{12}^q s_{23}^q s_{13}^q e^{i\delta^q} & c_{12}^q c_{23}^q - s_{12}^q s_{23}^q s_{13}^q e^{i\delta^q} & s_{23}^q c_{13}^q  \\ 
 s_{12}^q s_{23}^q - c_{12}^q c_{23}^q s_{13}^q e^{i\delta^q} &- c_{12}^q s_{23}^q - s_{12}^q c_{23}^q s_{13}^q e^{i\delta^q} & c_{23}^q c_{13}^q \\ 
\end{pmatrix} \; , 
\label{eq:VCKMparam}
\end{align}
where $\theta_{ij}^q$ ($i<j=1,2,3$) are the three quark mixing angles with $c_{i j}^q \equiv \cos \theta_{i j}^q$,~$s_{i j}^q \equiv \sin \theta_{i j}^q$, and $\delta^q$ is the CKM CP-violating phase.

In the limit $\mathbf{M}_D \ll \mathbf{M}_R$, the effective (light) neutrino mass matrix is given by the well-known seesaw formula: 
\begin{align}
\mathbf{M}_\nu=-\mathbf{M}_D \mathbf{M}_R^{-1} \mathbf{M}_D^T \; .
\label{eq:EffectiveNeutrino}
\end{align}
Charged-lepton and active-neutrino fields can be rotated to their physical basis as: 
\begin{equation}
\begin{aligned}
    e_{L,R} \rightarrow \mathbf{U}_{L,R}^e e_{L,R}& &\Rightarrow& &\mathbf{U}_L^{e \dagger} \mathbf{M}_e \mathbf{U}_R^e = \mathbf{D}_e = \text{diag}(m_e,m_\mu,m_\tau)& \; , 
    \\
    \nu_{L} \rightarrow \mathbf{U}^\nu \nu_{L}& &\Rightarrow& &\mathbf{U}^{\nu \dagger} \mathbf{M}_\nu \mathbf{U}^\nu = \mathbf{D}_\nu = \text{diag}(m_1,m_2,m_3)& \; ,
\label{eq:leptonOrdering}
\end{aligned}
\end{equation}
where $m_{e, \mu, \tau}$ and $m_{1,2,3}$ are the physical charged-lepton and effective neutrino masses, respectively. As only two RH neutrinos $\nu_R$ are included in our model, the lightest neutrino remains massless, i.e. $m_1=0$ ($m_3=0$) for NO (IO). For a given $\mathbf{M}_e$ and $\mathbf{M}_\nu$, the unitary matrices $\mathbf{U}_L^e$, $\mathbf{U}_R^e$ and $\mathbf{U}^\nu$ are determined through the diagonalization of the Hermitian matrices $\mathbf{H}_{e}=\mathbf{M}_{e} \mathbf{M}_{e}^{\dagger}$ and $\mathbf{H}_{e,\nu}^{\prime}=\mathbf{M}_{e,\nu}^{\dagger} \mathbf{M}_{e,\nu}$, as follows,
\begin{align}
\mathbf{U}_L^{e \dagger} \mathbf{H}_{e} \mathbf{U}_L^e = \mathbf{U}_R^{e \dagger} \mathbf{H}_{e}^{\prime} \mathbf{U}_R^e =\mathbf{D}_{e}^2\; ,  \quad \mathbf{U}^{\nu \dagger} \mathbf{H}_\nu^{\prime} \mathbf{U}^\nu=\mathbf{D}_\nu^2 \; .
\end{align}
The above procedure results in the lepton mixing matrix
\begin{align}
\mathbf{U}=\mathbf{U}_L^{e \dagger} \mathbf{U}^\nu \; ,
\end{align}
which, for the Majorana case with a massless neutrino, can be parameterized as~\cite{Rodejohann:2011vc}
\begin{align}
\U=\begin{pmatrix}
c_{12}^\ell c_{13}^\ell&s_{12}^\ell c_{13}^\ell&s_{13}^\ell\\
-s_{12}^\ell c_{23}^\ell-c_{12}^\ell s_{23}^\ell s_{13}^\ell e^{i\delta^\ell}&c_{12}^\ell c_{23}^\ell-s_{12}^\ell s_{23}^\ell s_{13}^\ell e^{i\delta^\ell}&s_{23}^\ell c_{13}^\ell e^{i\delta^\ell}\\
s_{12}^\ell s_{23}^\ell-c_{12}^\ell c_{23}^\ell s_{13}^\ell e^{i\delta^\ell}&-c_{12}^\ell s_{23}^\ell-s_{12}^\ell c_{23}^\ell s_{13}^\ell e^{i\delta^\ell}&c_{23}^\ell c_{13}^\ell e^{i\delta^\ell}
\end{pmatrix} \begin{pmatrix}
1&0&0\\
0&e^{i\frac{\alpha}{2}}&0\\
0&0&1
\end{pmatrix}\;,
\label{eq:UPMNSparam}
\end{align}
where $c_{ij}^\ell\equiv\cos\theta_{ij}^\ell$ and $s_{ij}^\ell\equiv\sin\theta_{ij}^\ell$ with $\theta_{ij}^\ell$ ($i<j=1,2,3$) being the lepton mixing angles, $\delta^\ell$ is the leptonic Dirac CP-violating phase and $\alpha$ is a Majorana phase.

\subsection{Strong CP problem}
\label{sec:strongCP}

Since the PQ symmetry is anomalous with respect to QCD, the transformation of Eq.~\eqref{eq:Lyuk2hdm} into the mass basis~\eqref{eq:Mass2hdm}, is reflected in the QCD topological term as,
\begin{align}
    \mathcal{L}_\theta = \left(\bar{\theta} + \frac{N a}{v_a}\right) \frac{g_s^2}{32 \pi^2} G^a_{\mu \nu} \tilde{G}^{\mu \nu}_a \; , \quad 
    N = \sum_{\alpha=1}^3 (2\chi_{q_\alpha}^L - \chi_{u_\alpha}^R - \chi_{d_\alpha}^R) \; ,
    \label{eq:Ncolor}
\end{align}
where the shift of the strong CP phase $\overline{\theta}$ is controlled by the color anomaly factor $N$. To implement the axion solution to the strong CP problem it is required that $N \neq 0$. Furthermore, from the above and Eq.~\eqref{eq:axionandvev}, we identify the axion decay constant as,
\begin{equation}
    f_a = \frac{f_{\text{PQ}}}{N} = \frac{v_{a}}{N} \; .
    \label{eq:axionfa}
\end{equation}

The axion becomes massive due to QCD instanton effects, which generate a non-perturbative potential that explicitly breaks the PQ symmetry, giving the axion a mass proportional to the QCD topological susceptibility. The up-to-date QCD axion mass at next-to-leading order (NLO) is~\cite{GrillidiCortona:2015jxo},
\begin{equation}
    m_a = 5.70(7) \left(\frac{10^{12} \text{GeV}}{f_a}\right) \mu \text{eV} \; .
    \label{eq:axionmass}
\end{equation}
This relation between $m_a$ and $f_a$ is a model-independent prediction of the QCD axion if the only explicit breaking of the PQ symmetry is by non-perturbative QCD effects.

\section{Minimal flavored Peccei-Quinn symmetries}
\label{sec:symmetries}

In general, the quark and lepton mass matrices are completely arbitrary, containing more independent parameters than those experimentally measured. Our goal is to identify the set of maximally-restrictive mass matrices that are both compatible with fermion data and realizable through the flavored PQ symmetries, taking as reference our previous work~\cite{Rocha:2024twm}. In this way, the same symmetry imposed to solve the strong CP problem, also addresses the observed fermion masses and mixing patterns.

A given set of texture-zero mass matrices is realizable through the PQ symmetry if it satisfies the following set of equations [see Eqs.~\eqref{eq:YukawaZeros} and~\eqref{Eq:MassMatrices}]:
\begin{align}
(\mathbf{M}_f)_{\alpha \beta} &= 0 \Leftrightarrow (\Theta^f_1)_{\alpha \beta} \neq 0 \  \text{(mod $2\pi$)} \ \wedge \ (\Theta^f_2)_{\alpha \beta} \neq 0 \ \text{(mod $2\pi$)} \; , \nonumber \\
(\mathbf{M}_D)_{\alpha j} &= 0 \Leftrightarrow (\Theta^D_1)_{\alpha j} \neq 0 \  \text{(mod $2\pi$)} \ \wedge \ (\Theta^D_2)_{\alpha j} \neq 0 \ \text{(mod $2\pi$)} \; , \nonumber \\
(\mathbf{M}_R)_{ij} &= 0 \Leftrightarrow (\Theta^R_{0})_{ij} \neq 0 \  \text{(mod $2\pi$)} \ \wedge \ (\Theta^R_{1})_{ij} \neq 0 \ \text{(mod $2\pi$)} \ \wedge \ (\Theta^R_{2})_{ij} \neq 0 \ \text{(mod $2\pi$)} \; .
\label{eq:canonicalchargesSystem}
\end{align}
To assess compatibility with quark, charged-lepton, and neutrino data within the $\nu$DFSZ framework [see Eq.~\eqref{Eq:MassMatrices}], we employ a standard $\chi^2$-analysis, with the function:
\begin{equation}
\chi^2(x) = \sum_i \frac{\left[\mathcal{P}_i(x) - \mathcal{O}_i\right]^2}{\sigma_i^2} \; ,
\label{eq:chi2}
\end{equation}
where $x$ denotes the input parameters, i.e., the matrix elements of $\mathbf{M}_{d}$, $\mathbf{M}_{u}$, $\mathbf{M}_{e}$, $\mathbf{M}_{D}$ and $\mathbf{M}_{R}$; $\mathcal{P}_i(x)$ is the model output for a given observable with best-fit value $\mathcal{O}_i$, and $\sigma_i$ denotes its $1\sigma$ experimental uncertainty. In our search for viable sets, we use the current data reported in Table~\ref{tab:data} and require that the $\chi^2$-function is minimized with respect to ten observables in the quark sector: the six quark masses $m_{d,s,b}$ and $m_{u,c,t}$, as well as the CKM parameters -- the three mixing angles $\theta_{12,23,13}^q$ and the CP-violating phase $\delta^q$; and nine observables in the lepton sector: three charged-lepton masses $m_{e,\mu,\tau}$, the two neutrino mass-squared differences $\Delta m^2_{21}, \Delta m^2_{31}$, and the lepton mixing matrix parameters -- the three mixing angles $\theta_{12,23,13}^\ell$ and the Dirac CP-violating phase $\delta^\ell$. Note that, for the neutrino oscillation parameters, $\chi^2(x)$ is computed using the one-dimensional profiles $\chi^2(\sin^2 \theta^\ell_{ij})$ and $\chi^2(\Delta m^2_{ij})$, and the two-dimensional~(2D) distribution $\chi^2(\delta^\ell,\sin^2 \theta^\ell_{23})$ for $\delta^\ell$ and $\theta_{23}^\ell$ given in Ref.~\cite{deSalas:2020pgw}. In our analysis, we consider a set of mass matrices to be compatible with data if the observables in Table~\ref{tab:data} fall within the $1\sigma$ range at the $\chi^2$-function minimum.

\begin{table}[t!]
    \renewcommand*{\arraystretch}{0.35}
    \begin{minipage}[b]{0.45\textwidth}
        \centering
            \begin{tabular}{|r|cc|}
                \cline{2-3}
                \multicolumn{1}{c|}{} & & \\
                \multicolumn{1}{c|}{} & \multicolumn{2}{c|}{Decomposition} \\[5pt]
                \hline
                & & \\ 
                Texture \; \; \; \; \; \; \; & $\mathbf{Y}_1^{d}$ & $\mathbf{Y}_2^{d}$ \\
                & & \\
                \hline
                & & \\ & & \\  
                \renewcommand{\arraystretch}{1.0} $4_{3}^{d}\sim \begin{pmatrix}
                0 & 0 & \times \\
                0 & \times & \times \\
                \times & \times & 0\\
                \end{pmatrix}$ \; \;    &    
                \renewcommand{\arraystretch}{1.0} $\begin{pmatrix}
                0 & 0 & \times \\
                0 & \times & 0 \\
                \times & 0 & 0 \\
                \end{pmatrix}$ &
                \renewcommand{\arraystretch}{1.0} $\begin{pmatrix}
                0 & 0 & 0 \\
                0 & 0 & \times \\
                0 & \times & 0 \\
                \end{pmatrix}$ 
                \\
                & & \\ & & \\
                \renewcommand{\arraystretch}{1.0} $5_{1}^{d}\sim \begin{pmatrix}
                0 & 0 & \times \\
                0 & \times & 0 \\
                \times & 0 & \times\\
                \end{pmatrix}$ \; \;   
                &   
                \renewcommand{\arraystretch}{1.0} $\begin{pmatrix}
                0 & 0 & \times \\
                0 & 0 & 0 \\
                \times & 0 & 0 \\
                \end{pmatrix}$ &
                \renewcommand{\arraystretch}{1.0} $\begin{pmatrix}
                0 & 0 & 0 \\
                0 & \times & 0 \\
                0 & 0 & \times \\
                \end{pmatrix}$
                \\
                & & \\ & & \\
                \hline
                & & \\ 
                 Texture  \; \; \; \; \; \; \; & $\mathbf{Y}_1^{u}$ & $\mathbf{Y}_2^{u}$ \\
                & & \\ 
                \hline
                & & \\ & & \\
                \renewcommand{\arraystretch}{1.0} $\mathbf{P}_{12}{5}_{1}^u\mathbf{P}_{23}\sim \begin{pmatrix}
                0 & 0 & \times \\
                0 & \bullet & 0 \\
                \times & \times & 0 \\
                \end{pmatrix}$ \; \; 
                &    
                \renewcommand{\arraystretch}{1.0} $\begin{pmatrix}
                0 & 0 & \times \\
                0 & 0 & 0 \\
                0 & \times & 0 \\
                \end{pmatrix}$ &
                \renewcommand{\arraystretch}{1.0} $\begin{pmatrix}
                0 & 0 & 0 \\
                0 & \bullet & 0 \\
                \times & 0 & 0 \\
                \end{pmatrix}$ 
                \\ & & \\ & & \\
                \renewcommand{\arraystretch}{1.0} $\mathbf{P}_{123}5_{1}^u \mathbf{P}_{12}\sim \begin{pmatrix}
                0 & \times & \bullet \\
                0 & 0 & \times \\
                \times & 0 & 0\\
                \end{pmatrix}$ \; \; 
                &    
                \renewcommand{\arraystretch}{1.0} $\begin{pmatrix}
                0 & \times & 0 \\
                0 & 0 & \times \\
                0 & 0 & 0 \\
                \end{pmatrix}$ &
                \renewcommand{\arraystretch}{1.0} $\begin{pmatrix}
                0 & 0 & \bullet \\
                0 & 0 & 0 \\
                \times & 0 & 0 \\
                \end{pmatrix}$ 
                \\ & & \\ & & \\
                \renewcommand{\arraystretch}{1.0} $\mathbf{P}_{12}{4}_{3}^u\sim \begin{pmatrix}
                0 & \bullet & \times \\
                0 & 0 & \times \\
                \times & \times & 0\\
                \end{pmatrix}$ \; \;     
                &    
                \renewcommand{\arraystretch}{1.0} $\begin{pmatrix}
                0 & 0 & \times \\
                0 & 0 & 0 \\
                0 & \times & 0 \\
                \end{pmatrix}$ &
                \renewcommand{\arraystretch}{1.0} $\begin{pmatrix}
                0 & \bullet & 0 \\
                0 & 0 & \times \\
                \times & 0 & 0 \\
                \end{pmatrix}$ 
                \\ & & \\ & & \\
                \renewcommand{\arraystretch}{1.0} $\mathbf{P}_{321}4_{3}^u\mathbf{P}_{23}\sim \begin{pmatrix}
                0 & \bullet & \times \\
                \times & 0 & \times \\
                0 & \times & 0\\
                \end{pmatrix}$ \; \; 
                &           
                \renewcommand{\arraystretch}{1.0} $\begin{pmatrix}
                0 & 0 & \times \\
                \times & 0 & 0 \\
                0 & \times & 0 \\
                \end{pmatrix}$ &
                \renewcommand{\arraystretch}{1.0} $\begin{pmatrix}
                0 & \bullet & 0 \\
                0 & 0 & \times \\
                0 & 0 & 0 \\
                \end{pmatrix}$ 
                \\ & & \\ & & \\
                \hline
                & & \\
                 Texture  \; \; \; \; \; \; \;  & $\mathbf{Y}_1^{e}$ & $\mathbf{Y}_2^{e}$ \\
                & & \\
                \hline
                & & \\ & & \\  
                \renewcommand{\arraystretch}{1.0} $4_{3}^{e}\sim \begin{pmatrix}
                0 & 0 & \times \\
                0 & \times & \times \\
                \times & \times & 0\\
                \end{pmatrix}$ \; \;    &    
                \renewcommand{\arraystretch}{1.0} $\begin{pmatrix}
                0 & 0 & \times \\
                0 & \times & 0 \\
                \times & 0 & 0 \\
                \end{pmatrix}$ &
                \renewcommand{\arraystretch}{1.0} $\begin{pmatrix}
                0 & 0 & 0 \\
                0 & 0 & \times \\
                0 & \times & 0 \\
                \end{pmatrix}$ 
                \\
                & & \\ & & \\
                \renewcommand{\arraystretch}{1.0} $5_{1}^{e}\sim \begin{pmatrix}
                0 & 0 & \times \\
                0 & \times & 0 \\
                \times & 0 & \times\\
                \end{pmatrix}$ \; \;   
                &   
                \renewcommand{\arraystretch}{1.0} $\begin{pmatrix}
                0 & 0 & \times \\
                0 & 0 & 0 \\
                \times & 0 & 0 \\
                \end{pmatrix}$ &
                \renewcommand{\arraystretch}{1.0} $\begin{pmatrix}
                0 & 0 & 0 \\
                0 & \times & 0 \\
                0 & 0 & \times \\
                \end{pmatrix}$
                \\
                & & \\ & & \\
                \hline
            \end{tabular}
    \end{minipage}%
    \hspace{1.0mm}
    \begin{minipage}[b]{0.5\textwidth}
        \centering
        \begin{tabular}{|r|cccccc|}
        \cline{2-7}
        \multicolumn{1}{c|}{} & & & & & &\\
        \multicolumn{1}{c|}{} & \multicolumn{6}{c|}{Decomposition} \\[5pt]
        \hline
        &  &  & &  &  &  \\ 
         Texture \;\;\;\;\;\; &   \multicolumn{3}{c}{\; \; \; \;  $\Y_1^D$} & \multicolumn{3}{c|}{  $\Y_2^D$ }    \\ 
        &  &  & &  &  &  \\ 
        \hline
        &  &  & &  &  &  \\ &  &  & &  &  &  \\ 
            \renewcommand{\arraystretch}{1.0} $2_{1}^{D}\sim \begin{pmatrix}
            \bullet & 0 \\
            \bullet & \bullet \\
            0 & \bullet \\
            \end{pmatrix}$ \; \; 
            & \multicolumn{3}{c}{\; \; \;  \renewcommand{\arraystretch}{1.0} $ \begin{pmatrix}
            0 & 0  \\
            \bullet & 0  \\
            0 & \bullet  \\
            \end{pmatrix}$ }
            & \multicolumn{3}{c|}{\ \renewcommand{\arraystretch}{1.0} $ \begin{pmatrix}
            \bullet & 0  \\
            0 & \bullet  \\
            0 & 0  \\
            \end{pmatrix}$ \; } \\
        &  &  & &  &  &  \\ &  &  & &  &  &  \\ 
            \renewcommand{\arraystretch}{1.0} $2_{2}^{D} \sim 
            \begin{pmatrix}
            \bullet & \bullet \\
            \bullet & 0 \\
            0 & \bullet \\
            \end{pmatrix}$ \; \; 
            & \multicolumn{3}{c}{\; \; \; \renewcommand{\arraystretch}{1.0} $ \begin{pmatrix}
            \bullet & 0  \\
            0 & 0  \\
            0 & \bullet  \\
            \end{pmatrix}$ }
            & \multicolumn{3}{c|}{\renewcommand{\arraystretch}{1.0} $ \begin{pmatrix}
            0 & \bullet  \\
            \bullet & 0  \\
            0 & 0  \\
            \end{pmatrix}$\;} \\
        &  &  & &  &  &  \\ &  &  & &  &  &  \\ 
            \renewcommand{\arraystretch}{1.0} $3_{1}^{D}\sim \begin{pmatrix}
            \bullet & \bullet \\
            \bullet & 0 \\
            0 & 0 \\
            \end{pmatrix}$ \; \; 
            & \multicolumn{3}{c}{\; \; \; \renewcommand{\arraystretch}{1.0} $ \begin{pmatrix}
            0 & \bullet  \\
            \bullet & 0  \\
            0 & 0  \\
            \end{pmatrix}$ }
            & \multicolumn{3}{c|}{\renewcommand{\arraystretch}{1.0} $ \begin{pmatrix}
            \bullet & 0  \\
            0 & 0  \\
            0 & 0  \\
            \end{pmatrix}$\;} \\
        &  &  & &  &  &  \\ &  &  & &  &  &  \\ 
            \renewcommand{\arraystretch}{1.0}$3_{2}^{D}\sim \begin{pmatrix}
            \bullet & 0 \\
            \bullet & 0 \\
            0 & \bullet \\
            \end{pmatrix}$ \; \; 
            & \multicolumn{3}{c}{\; \; \; \renewcommand{\arraystretch}{1.0} $ \begin{pmatrix}
            0 & 0  \\
            \bullet & 0  \\
            0 & 0  \\
            \end{pmatrix}$ }
            & \multicolumn{3}{c|}{\renewcommand{\arraystretch}{1.0} $ \begin{pmatrix}
            \bullet & 0  \\
            0 & 0  \\
            0 & \bullet  \\
            \end{pmatrix}$ \;} \\
        &  &  & &  &  &  \\ &  &  & &  &  &  \\ 
            \renewcommand{\arraystretch}{1.0} $3_{3}^{D}\sim \begin{pmatrix}
            \bullet & 0 \\
            0 & \bullet \\
            0 & \bullet \\
            \end{pmatrix}$ \; \; 
            & \multicolumn{3}{c}{\; \; \; \renewcommand{\arraystretch}{1.0} $ \begin{pmatrix}
            \bullet & 0  \\
            0 & 0  \\
            0 & \bullet  \\
            \end{pmatrix}$ }
            & \multicolumn{3}{c|}{\renewcommand{\arraystretch}{1.0} $ \begin{pmatrix}
            0 & 0  \\
            0 & \bullet  \\
            0 & 0  \\
            \end{pmatrix}$ \;}\\
        &  &  & &  &  &  \\ &  &  & &  &  &  \\ 
            \renewcommand{\arraystretch}{1.0} $3_{4}^{D}\sim \begin{pmatrix}
            0 & 0 \\
            \bullet & 0 \\
            \bullet & \bullet \\
            \end{pmatrix}$ \; \; 
            & \multicolumn{3}{c}{\; \; \; \renewcommand{\arraystretch}{1.0} $ \begin{pmatrix}
            0 & 0  \\
            0 & 0  \\
            \bullet & 0  \\
            \end{pmatrix}$ }
            & \multicolumn{3}{c|}{\renewcommand{\arraystretch}{1.0} $ \begin{pmatrix}
            0 & 0  \\
            \bullet & 0  \\
            0 & \bullet  \\
            \end{pmatrix}$\;} \\
        &  &  & &  &  &  \\ &  &  & &  &  &  \\ 
        \hline
        &  &  & &  &  &  \\        
        Texture \;\;\;\;\;\; & \multicolumn{2}{c}{$\mathbf{M}^0_R$} & \multicolumn{2}{c}{$\mathbf{Y}^1_R$} & \multicolumn{2}{c|}{$\mathbf{Y}^2_R$} \\
        &  &  & &  &  &  \\
        \hline
        &  &  & &  &  &  \\ &  &  & &  &  &  \\ 
        \renewcommand{\arraystretch}{1.0} $1_{1,1}^{R}\sim \begin{pmatrix}
            0 & \times \\
            \cdot & \times \\
            \end{pmatrix}$ \; \;  
            & \multicolumn{2}{c}{\renewcommand{\arraystretch}{1.0} $\begin{pmatrix}
            0 & 0  \\
            \cdot & \times  \\
            \end{pmatrix}$}
            & \multicolumn{2}{c}{\renewcommand{\arraystretch}{1.0} $\begin{pmatrix}
            0 & \times  \\
            \cdot & 0  \\
            \end{pmatrix}$ }
            & \multicolumn{2}{c|}{\renewcommand{\arraystretch}{1.0} $\begin{pmatrix}
            0 & 0  \\
            \cdot & 0  \\
            \end{pmatrix}$} \\
        &  &  & &  &  &  \\ &  &  & &  &  &  \\ 
        \renewcommand{\arraystretch}{1.0} $1_{1,2}^{R}\sim \begin{pmatrix}
            0 & \times \\
            \cdot & \times \\
            \end{pmatrix}$ \; \;  
            & \multicolumn{2}{c}{\renewcommand{\arraystretch}{1.0} $\begin{pmatrix}
            0 & 0  \\
            \cdot & 0  \\
            \end{pmatrix}$}
            & \multicolumn{2}{c}{\renewcommand{\arraystretch}{1.0} $\begin{pmatrix}
            0 & \times  \\
            \cdot & 0  \\
            \end{pmatrix}$ }
            & \multicolumn{2}{c|}{\renewcommand{\arraystretch}{1.0} $\begin{pmatrix}
            0 & 0  \\
            \cdot & \times  \\
            \end{pmatrix}$} \\
        &  &  & &  &  &  \\ &  &  & &  &  &  \\ 
        \renewcommand{\arraystretch}{1.0} $1_{1,3}^{R}\sim \begin{pmatrix}
            0 & \times \\
            \cdot & \times \\
            \end{pmatrix}$ \; \;  
            &\multicolumn{2}{c}{ \renewcommand{\arraystretch}{1.0} $\begin{pmatrix}
            0 & 0  \\
            \cdot & 0  \\
            \end{pmatrix}$}
            & \multicolumn{2}{c}{\renewcommand{\arraystretch}{1.0} $\begin{pmatrix}
            0 & 0  \\
            \cdot & \times  \\
            \end{pmatrix}$ }
            & \multicolumn{2}{c|}{\renewcommand{\arraystretch}{1.0} $\begin{pmatrix}
            0 & \times  \\
            \cdot & 0  \\
            \end{pmatrix}$} \\
        &  &  & &  &  &  \\ &  &  & &  &  &  \\ 
        \renewcommand{\arraystretch}{1.0} $1_{1,4}^{R}\sim \begin{pmatrix}
            0 & \times \\
            \cdot & \times \\
            \end{pmatrix}$ \; \;  
            & \multicolumn{2}{c}{\renewcommand{\arraystretch}{1.0} $\begin{pmatrix}
            0 & 0  \\
            \cdot & \times  \\
            \end{pmatrix}$}
            & \multicolumn{2}{c}{\renewcommand{\arraystretch}{1.0} $\begin{pmatrix}
            0 & 0  \\
            \cdot & 0  \\
            \end{pmatrix}$ }
            & \multicolumn{2}{c|}{\renewcommand{\arraystretch}{1.0} $\begin{pmatrix}
            0 & \times  \\
            \cdot & 0  \\
            \end{pmatrix}$} \\
        &  &  & &  &  &  \\ &  &  & &  &  &  \\ 
        \renewcommand{\arraystretch}{1.0} $1_{2}^{R}\sim \begin{pmatrix}
            \times & \times \\
            \cdot & 0 \\
            \end{pmatrix}$ \; \;
            & \multicolumn{2}{c}{\renewcommand{\arraystretch}{1.0} $\begin{pmatrix}
            0 & 0  \\
            \cdot & 0  \\
            \end{pmatrix}$ }
            & \multicolumn{2}{c}{\renewcommand{\arraystretch}{1.0} $\begin{pmatrix}
            0 & \times  \\
            \cdot & 0  \\
            \end{pmatrix}$ }
            & \multicolumn{2}{c|}{\renewcommand{\arraystretch}{1.0} $\begin{pmatrix}
            \times & 0  \\
            \cdot & 0  \\
            \end{pmatrix}$} \\
        &  &  & &  &  &  \\ &  &  & &  &  &  \\ 
        \renewcommand{\arraystretch}{1.0} $2_{1}^{R}\sim \begin{pmatrix}
            0 & \times \\
            \cdot & 0 \\
            \end{pmatrix}$ \; \;    
            & \multicolumn{2}{c}{\renewcommand{\arraystretch}{1.0} $\begin{pmatrix}
            0 & \times \\
            \cdot & 0 \\
            \end{pmatrix}$ }
            & \multicolumn{2}{c}{\renewcommand{\arraystretch}{1.0} $\begin{pmatrix}
            0 & 0  \\
            \cdot & 0  \\
            \end{pmatrix}$ }
            & \multicolumn{2}{c|}{\renewcommand{\arraystretch}{1.0} $\begin{pmatrix}
            0 & 0  \\
            \cdot & 0  \\
            \end{pmatrix}$} \\
        &  &  & &  &  &  \\ &  &  & &  &  &  \\ 
        \hline
        \end{tabular}
    \end{minipage}
    \caption{Realizable decomposition into Yukawa matrices of the quark and charged-lepton (left) and neutrino (right) mass matrices for the texture pairs of Tables~\ref{tab:quarkcharges} and~\ref{tab:leptoncharges}. A matrix entry ``$0$" denotes a texture zero, ``$\times$" and~``$\bullet$" are a real positive and complex entry, respectively. The symmetric character of the Majorana matrix is marked by a ``$\cdot$".} 
    \label{tab:Matrices}
\end{table}
\begin{table}[t!]
    \centering
    \renewcommand*{\arraystretch}{1.5}
    \begin{tabular}{|c|c|ccc|c|}
            \hline
            \; Model \; &
            ($\mathbf{M}_d$,$\mathbf{M}_u)$  &
            $\chi^L_{q_\alpha} + \frac{s_\beta^2}{3} $ &
            $\chi^R_{d_\alpha} - \frac{2 s_\beta^2}{3}$
            & $\chi^R_{u_\alpha} + \frac{4 s_\beta^2}{3}$ & $\;\; N \;\;$
            \\ \hline
            $\text{Q}_\text{1}^\text{I}$ &
            \multirow{2}{*}{$(4_{3}^d,\mathbf{P}_{12}{5}_{1}^u\mathbf{P}_{23})$}              &      $\left(0,1,2 \right)$    &    $\left(2,1,0 \right)$  &  $\left(3,2,0 \right)$   &  $\;\;-2\;\;$ 
            \\
            $\text{Q}_\text{1}^\text{II}$ &
                       &      $\left(0,-1,-2 \right)$    &    $\left(-3,-2,-1 \right)$  &  $\left(-2,-1,1 \right)$   &  $\;\; 2\;\;$ 
            \\
            \hline
            $\text{Q}_\text{2}^\text{I}$ &
            \multirow{2}{*}{$(4_{3}^d,\mathbf{P}_{123}5_{1}^u \mathbf{P}_{12})$}             &      $\left(0,1,2 \right)$    &    $\left(2,1,0 \right)$  &  $\left(3,0,1 \right)$  &  $\;\;-1\;\;$ 
            \\
            $\text{Q}_\text{2}^\text{II}$ &
                          &      $\left(0,-1,-2 \right)$    &    $\left(-3,-2,-1 \right)$  &  $\left(-2,1,0 \right)$  &  $\;\; 1\;\;$ 
            \\
            \hline
            $\text{Q}_\text{3}^\text{I}$ &
            \multirow{2}{*}{$(5_{1}^d,\mathbf{P}_{12}{4}_{3}^u)$}     &      $\left(0,-1,1 \right)$    &    $\left(1,-2,0 \right)$  &  $\left(2,1,0 \right)$ &     $\;\;-2\;\;$
            \\
            $\text{Q}_\text{3}^\text{II}$ &
                 &      $\left(0,1,-1 \right)$    &    $\left(-2,1,-1 \right)$  &  $\left(-1,0,1 \right)$ &     $\;\;2\;\;$
            \\
            \hline
            $\text{Q}_\text{4}^\text{I}$ &
             \multirow{2}{*}{$(5_{1}^d,\mathbf{P}_{321}4_{3}^u\mathbf{P}_{23})$}         &      $\left(0,-1,1\right)$    &    $\left(1,-2,0 \right)$  &  $\left(-1,1,0 \right)$ & $\;\;1\;\;$ \\
            $\text{Q}_\text{4}^\text{II}$ &
                    &      $\left(0,1,-1\right)$    &    $\left(-2,1,-1 \right)$  &  $\left(2,0,1 \right)$ & $\;\;-1\;\;$ \\
            \hline 
        \end{tabular}
    \caption{
        Maximally-restrictive quark models ${\rm Q}^{\text{I,II}}_{1-4}$ made out of matrix pairs ($\mathbf{M}_d$,$\mathbf{M}_u$) compatible with data at~$1 \sigma$ CL (see Table~\ref{tab:data}). The superscript ``I" stands for the Yukawa decompositions given in Table~\ref{tab:Matrices}, while for models with superscript``II" the Yukawa textures are switched $\mathbf{Y}_1^{u,d} \leftrightarrow \mathbf{Y}_2^{u,d}$.
        $N$ is the color anomaly factor [see Eq.~\eqref{eq:Ncolor}]. Quark field charges $\chi_{q_\alpha}^L$, $\chi_{d_\alpha}^R$ and $\chi_{u_\alpha}^R$, follow the notation of Eq.~\eqref{eq:PQSym}.}
    \label{tab:quarkcharges}
\end{table}
\begin{table}[t!]
\centering
\renewcommand{\arraystretch}{1.5}
\begin{tabular}{|c|c|cccc|}
      \hline
      \; Model \; & ($\mathbf{M}_{e}$,$\mathbf{M}_D$,$\mathbf{M}_R$) &
      $\chi_\sigma$ &
      $\chi^L_{\ell_\alpha} - s_\beta^2$ &
      $\chi^R_{e_\alpha} - 2s_\beta^2$ &
      $\chi^R_{\nu_j}$ 
      \\ 
      \hline
      \renewcommand{\arraystretch}{1.0}\begin{tabular}[c]{@{}c@{}} $ $ \\ $\text{L}_\text{1}$\\ $ $\end{tabular} &
      \renewcommand{\arraystretch}{1.0}\begin{tabular}[c]{@{}c@{}} $ $ \\ $(4_{3}^{e},2_{1}^{D},2_{1}^{R})$\\ $ $\end{tabular} &
      \renewcommand{\arraystretch}{1.0}\begin{tabular}[c]{@{}c@{}}$ $\\ $1/2$\\ $ $\end{tabular} &
      \renewcommand{\arraystretch}{1.0}\begin{tabular}[c]{@{}c@{}}$ $\\ $\left(-3/2,-1/2,1/2 \right)$\\ $ $\end{tabular} &
      \renewcommand{\arraystretch}{1.0}\begin{tabular}[c]{@{}c@{}}$ $\\ $\left(1/2,-1/2,-3/2 \right)$\\ $ $\end{tabular} &
      \renewcommand{\arraystretch}{1.0}\begin{tabular}[c]{@{}c@{}}$ $\\ $\left(-1/2,1/2 \right)$\\ $ $\end{tabular} 
      \\
      \hline
      \renewcommand{\arraystretch}{1.4}\begin{tabular}[c]{@{}c@{}}$\text{L}_\text{2}$\\ $\text{L}_\text{3}$\end{tabular} &
      \renewcommand{\arraystretch}{1.4}\begin{tabular}[c]{@{}c@{}}$(4_{3}^{e},3_{1}^{D},1_{1,1}^{R})$\\ $(4_{3}^{e},3_{1}^{D},1_{1,2}^{R}) $\end{tabular} &
      \renewcommand{\arraystretch}{1.4}\begin{tabular}[c]{@{}c@{}} $1$\\ $1/2$\end{tabular} &
      \renewcommand{\arraystretch}{1.4}\begin{tabular}[c]{@{}c@{}} $\left(0,1,2 \right)$\\ $\left(-1/4,3/4,7/4\right)$\end{tabular} &
      \renewcommand{\arraystretch}{1.4}\begin{tabular}[c]{@{}c@{}} $\left(2 ,1,0 \right)$\\ $\left({7}/{4},{3}/{4},-{1}/{4}\right)$\end{tabular} &
      \renewcommand{\arraystretch}{1.4}\begin{tabular}[c]{@{}c@{}} $\left(1,0 \right)$\\ $\left({3}/{4},-{1}/{4} \right)$\end{tabular}
      \\ 
      \hline
      \renewcommand{\arraystretch}{1.0}\begin{tabular}[c]{@{}c@{}}$ $\\ $\text{L}_\text{4}$\\ $ $\end{tabular} &
      \renewcommand{\arraystretch}{1.0}\begin{tabular}[c]{@{}c@{}}$ $\\ $(4_{3}^{e},3_{2}^{D},1_{2}^{R})$\\ $ $\end{tabular} &
      \renewcommand{\arraystretch}{1.0}\begin{tabular}[c]{@{}c@{}}$ $\\ $1$\\ $ $\end{tabular} &
      \renewcommand{\arraystretch}{1.0}\begin{tabular}[c]{@{}c@{}}$ $\\ $\left(-{3}/{2},-{1}/{2},{1}/{2}  \right)$\\ $ $\end{tabular} &
      \renewcommand{\arraystretch}{1.0}\begin{tabular}[c]{@{}c@{}}$ $\\ $\left( {1}/{2},-{1}/{2},-{3}/{2} \right)$\\ $ $\end{tabular} &
      \renewcommand{\arraystretch}{1.0}\begin{tabular}[c]{@{}c@{}}$ $\\ $\left(-{1}/{2},{3}/{2} \right)$\\ $ $\end{tabular}
      \\ 
      \hline
      \renewcommand{\arraystretch}{1.0}\begin{tabular}[c]{@{}c@{}}$ $\\ $\text{L}_\text{5}$\\ $ $\end{tabular} &
      \renewcommand{\arraystretch}{1.0}\begin{tabular}[c]{@{}c@{}}$ $\\ $(4_{3}^{e},3_{2}^{D},1_{1,3}^{R})$\\ $ $\end{tabular} &
      \renewcommand{\arraystretch}{1.0}\begin{tabular}[c]{@{}c@{}}$ $\\ $1$\\ $ $\end{tabular} &
      \renewcommand{\arraystretch}{1.0}\begin{tabular}[c]{@{}c@{}}$ $\\ $\left(-{5}/{2},-{3}/{2},-{1}/{2}   \right)$\\ $ $\end{tabular} &
      \renewcommand{\arraystretch}{1.0}\begin{tabular}[c]{@{}c@{}}$ $\\ $\left(-{1}/{2} ,-{3}/{2},-{5}/{2} \right)$\\ $ $\end{tabular} &
      \renewcommand{\arraystretch}{1.0}\begin{tabular}[c]{@{}c@{}}$ $\\ $\left(-{3}/{2},{1}/{2} \right)$\\ $ $\end{tabular} 
      \\ 
      \hline
      \renewcommand{\arraystretch}{1.0}\begin{tabular}[c]{@{}c@{}}$ $\\ $\text{L}_\text{6}$ \\ $ $ \end{tabular} &
      \renewcommand{\arraystretch}{1.0}\begin{tabular}[c]{@{}c@{}}$ $\\ $(4_{3}^{e},3_{3}^{D},1_{1,3}^{R})$ \\ $ $ \end{tabular} &
      \renewcommand{\arraystretch}{1.0} \begin{tabular}[c]{@{}c@{}}$ $\\ $1$\\ $ $\end{tabular} &
      \renewcommand{\arraystretch}{1.0}\begin{tabular}[c]{@{}c@{}}$ $\\ $\left(-{3}/{2},-{1}/{2} ,{1}/{2} \right)$\\ $ $\end{tabular} &
      \renewcommand{\arraystretch}{1.0}\begin{tabular}[c]{@{}c@{}}$ $\\ $\left({1}/{2},-{1}/{2},-{3}/{2}  \right)$\\ $ $\end{tabular} &
      \renewcommand{\arraystretch}{1.0} \begin{tabular}[c]{@{}c@{}}$ $\\ $\left(-{3}/{2},{1}/{2} \right)$\\ $ $\end{tabular} 
      \\ 
      \hline
      \renewcommand{\arraystretch}{1.0}\begin{tabular}[c]{@{}c@{}}$ $\\ $\text{L}_\text{7}$\\ $ $\end{tabular} &
      \renewcommand{\arraystretch}{1.0}\begin{tabular}[c]{@{}c@{}}$ $\\ $(4_{3}^{e},3_{3}^{D},1_{2}^{R})$\\ $ $\end{tabular} &
      \renewcommand{\arraystretch}{1.0}\begin{tabular}[c]{@{}c@{}}$ $\\ $1$\\ $ $\end{tabular} &
      \renewcommand{\arraystretch}{1.0}\begin{tabular}[c]{@{}c@{}}$ $\\ $\left(-{1}/{2},{1}/{2},{3}/{2} \right)$\\ $ $\end{tabular} &
      \renewcommand{\arraystretch}{1.0}\begin{tabular}[c]{@{}c@{}}$ $\\ $\left({3}/{2},{1}/{2},-{1}/{2}  \right)$\\ $ $\end{tabular} &
      \renewcommand{\arraystretch}{1.0}\begin{tabular}[c]{@{}c@{}}$ $\\ $\left(-{1}/{2},{3}/{2} \right)$\\ $ $\end{tabular} 
      \\
      \hline
      \renewcommand{\arraystretch}{1.4}  \begin{tabular}[c]{@{}c@{}} $\text{L}_\text{8}$\\ $\text{L}_\text{9}$ \end{tabular} &
      \renewcommand{\arraystretch}{1.4}  \begin{tabular}[c]{@{}c@{}} $(4_{3}^{e},3_{4}^{D},1_{1,4}^{R})$\\ $(4_{3}^{e},3_{4}^{D},1_{1,3}^{R})$ \end{tabular} &
      \renewcommand{\arraystretch}{1.4}\begin{tabular}[c]{@{}c@{}} $1$\\ ${1}/{2}$\end{tabular} &
      \renewcommand{\arraystretch}{1.4}\begin{tabular}[c]{@{}c@{}} $\left(-3,-2,-1\right)$\\ $\left(-{11}/{4},-{7}/{4},-{3}/{4} \right)$\end{tabular} &
      \renewcommand{\arraystretch}{1.4}\begin{tabular}[c]{@{}c@{}} $\left(-1,-2,-3 \right)$\\ $\left(-{3}/{4},-{7}/{4},-{11}/{4} \right)$\end{tabular} &
      \renewcommand{\arraystretch}{1.4}\begin{tabular}[c]{@{}c@{}} $\left(-1,0 \right)$\\ $\left(-{3}/{4},{1}/{4} \right)$\end{tabular} 
      \\ 
      \hline
      \renewcommand{\arraystretch}{1.4}\begin{tabular}[c]{@{}c@{}} $\text{L}_\text{10}$ \\ $\text{L}_\text{11}$ \end{tabular} &
      \renewcommand{\arraystretch}{1.4}\begin{tabular}[c]{@{}c@{}} $(5_{1}^{e},2_{2}^{D},1_{1,4}^{R})$\\ $(5_{1}^{e},2_{2}^{D},1_{1,3}^{R})$\end{tabular} &
      \renewcommand{\arraystretch}{1.4}\begin{tabular}[c]{@{}c@{}} $1$\\ ${1}/{2}$\end{tabular} &
      \renewcommand{\arraystretch}{1.4}\begin{tabular}[c]{@{}c@{}} $\left(-1,-2,0 \right)$\\ $\left(-{3}/{4},-{7}/{4},{1}/{4} \right)$\end{tabular} &
      \renewcommand{\arraystretch}{1.4}\begin{tabular}[c]{@{}c@{}} $\left(0,-3, -1 \right)$\\ $\left({1}/{4},-{11}/{4},-{3}/{4} \right)$\end{tabular} &
      \renewcommand{\arraystretch}{1.4}\begin{tabular}[c]{@{}c@{}} $\left(-1,0 \right)$\\ $\left(-{3}/{4},{1}/{4} \right)$\end{tabular} 
      \\ 
      \hline
\end{tabular}
\caption{Maximally-restrictive lepton models ${\rm L}_{1-11}$ made out of the matrix sets ($\mathbf{M}_e$,$\mathbf{M}_D$,$\mathbf{M}_R$) compatible with data at~$1 \sigma$ CL for both NO and IO (see Table~\ref{tab:data}). The lepton field charges $\chi_{\ell_\alpha}^L$, $\chi_{e_\alpha}^R$ and $\chi_{\nu_j}^R$ follow the notation of Eq.~\eqref{eq:PQSym}.} 
\label{tab:leptoncharges}
\end{table}
Our search for maximally restrictive matrix sets is simplified by considering only physically equivalent sets, known as equivalence classes~\cite{Ludl:2014axa,Ludl:2015lta,GonzalezFelipe:2014zjk,Correia:2019vbn,Camara:2020efq,Rocha:2024twm}. In particular, any set of mass matrices belongs to the same equivalence class if they can be transformed into one another through the following weak-basis permutations:
\begin{align}
    \mathbf{M}_d \rightarrow \mathbf{P}_q^T \mathbf{M}_d \mathbf{P}_d\; , \;  \mathbf{M}_u \rightarrow \mathbf{P}_q^T \mathbf{M}_u \mathbf{P}_u \; , \; \mathbf{M}_e \rightarrow \mathbf{P}_\ell^T \mathbf{M}_e \mathbf{P}_e\; , \; \mathbf{M}_D \rightarrow \mathbf{P}_\ell^T \mathbf{M}_D \mathbf{P}_\nu\; , 
    \mathbf{M}_R \rightarrow \mathbf{P}_\nu^T \mathbf{M}_R \mathbf{P}_\nu \; ,
    \;
    \label{eq:Permutations}
\end{align}
where $\mathbf{P}_{\nu}$ are $2 \times 2$ permutation matrices, while the remaining $\mathbf{P}$'s are $3 \times 3$ permutation matrices.  Thus, we restrict our search for quark matrix sets to the equivalence classes derived in Ref.~\cite{Ludl:2014axa}. For the lepton sector, we adopt the charged-lepton equivalence classes from Ref~\cite{Ludl:2015lta}, where every possible permutation of $\mathbf{M}_e$ via $\mathbf{P}_\ell^T$ and $\mathbf{P}_e$ is considered. Additionally, $\mathbf{M}_D$ matrices related by column permutations belong to the same class. Therefore, we consider only one representative from each class. With this approach, for $\mathbf{M}_R$ we must account for all possible textures.

Our results for the maximally-restrictive sets $\left(\mathbf{M}_{d},\mathbf{M}_{u},\mathbf{M}_{e},\mathbf{M}_{D},\mathbf{M}_{R}\right)$ are summarized in Table~\ref{tab:Matrices}, where we present the quark and lepton mass matrix textures (leftmost column) and their corresponding decompositions (rightmost columns) in terms of the original Yukawa and mass matrices of the Lagrangian -- see Eqs.~\eqref{eq:Mass2hdm} and \eqref{Eq:MassMatrices}. All together, the number of independent parameters in the maximally-restrictive sets of fermion mass matrices matches the number of observables (masses and mixing parameters) in the quark and lepton sectors, meaning that all mass matrix elements are determined by data. Consequently, all non-vanishing Yukawa couplings are known up to a $s_\beta$ or $c_\beta$ factor. 

The PQ charges that realize the decompositions of Table~\ref{tab:Matrices} are listed in Table~\ref{tab:quarkcharges} and~\ref{tab:leptoncharges}. A few comments are in order: 
\begin{itemize}

\item \textbf{Minimal flavor patterns for quarks:} The maximally-restrictive quark textures realizable by a U$(1)$ Abelian flavor symmetry were found in Ref.~\cite{Rocha:2024twm}, where, without loss of generality, $\chi_1=0$, $\chi_2=1$, and $\chi^L_{q_1}=0$. To promote this framework to a flavored PQ scenario, i.e., to comply with Eq.~\eqref{eq:PQsymmetry}, we shift the flavor charges by $-2Y s_\beta^2$, where $Y$ is the hypercharge of the corresponding field~\footnote{This is possible due to the freedom to redefine the PQ charges through an admixture of the colour-anomaly-free global hypercharge and baryon number.}. Besides the models presented in Ref.~\cite{Rocha:2024twm}, which we label with the superscript ``I'', here we also consider those with $\mathbf{Y}_1^{u,d} \leftrightarrow \mathbf{Y}_2^{u,d}$, marked with ``II'', by redefining the PQ quark charges as $\chi^{L,R} \to -\chi^{L,R} + 2 Y (1+2\chi_1) - 2Y_{q_L}$. This ensures that $\chi^L_{q_1} = -2Y_{q_{L_1}} s_\beta^2$, as in the ``I'' cases. Note that, the I/II Yukawa ordering has physical implications regarding the predictions for the axion-to-photon couplings as shown in Sec.~\ref{sec:axionphoton}.

The PQ charge assignments of the maximally-restrictive quark models, along with the corresponding anomaly factor, are presented in Table~\ref{tab:quarkcharges}. The maximally restrictive matrix structures are shown in Table~\ref{tab:Matrices}.

\clearpage

\item \textbf{Minimal flavor patterns for leptons:} The maximally-restrictive matrix pairs $(\mathbf{M}_{e}, \mathbf{M}_D, \mathbf{M}_R)$, are identified by examining the equivalence classes with the largest number of texture zeros. We then check whether they can be realized by PQ symmetries solving Eqs.~\eqref{eq:canonicalchargesSystem}, and whether they are compatible with the current lepton data given in Table~\ref{tab:data}. If none of the equivalence classes under consideration passes this test, we add a non-zero entry to the mass matrix pairs and repeat the process until compatibility is achieved. This methodology, which is similar to that of previous works~\cite{GonzalezFelipe:2016tkv,Correia:2019vbn,Camara:2020efq,Rocha:2024twm}, led to eleven different models. The PQ charges~\footnote{Note that performing Yukawa permutations in both the quark and lepton sectors would lead to a physically equivalent scenario as the non-permuted case. Since we have considered permutations in the quark sector, we do not consider them in the lepton sector.} for each model are shown in Table~\ref{tab:leptoncharges} (their respective textures can be found in Table~\ref{tab:Matrices}).

\begin{figure}[!t]
    \centering
      \includegraphics[scale=0.85]{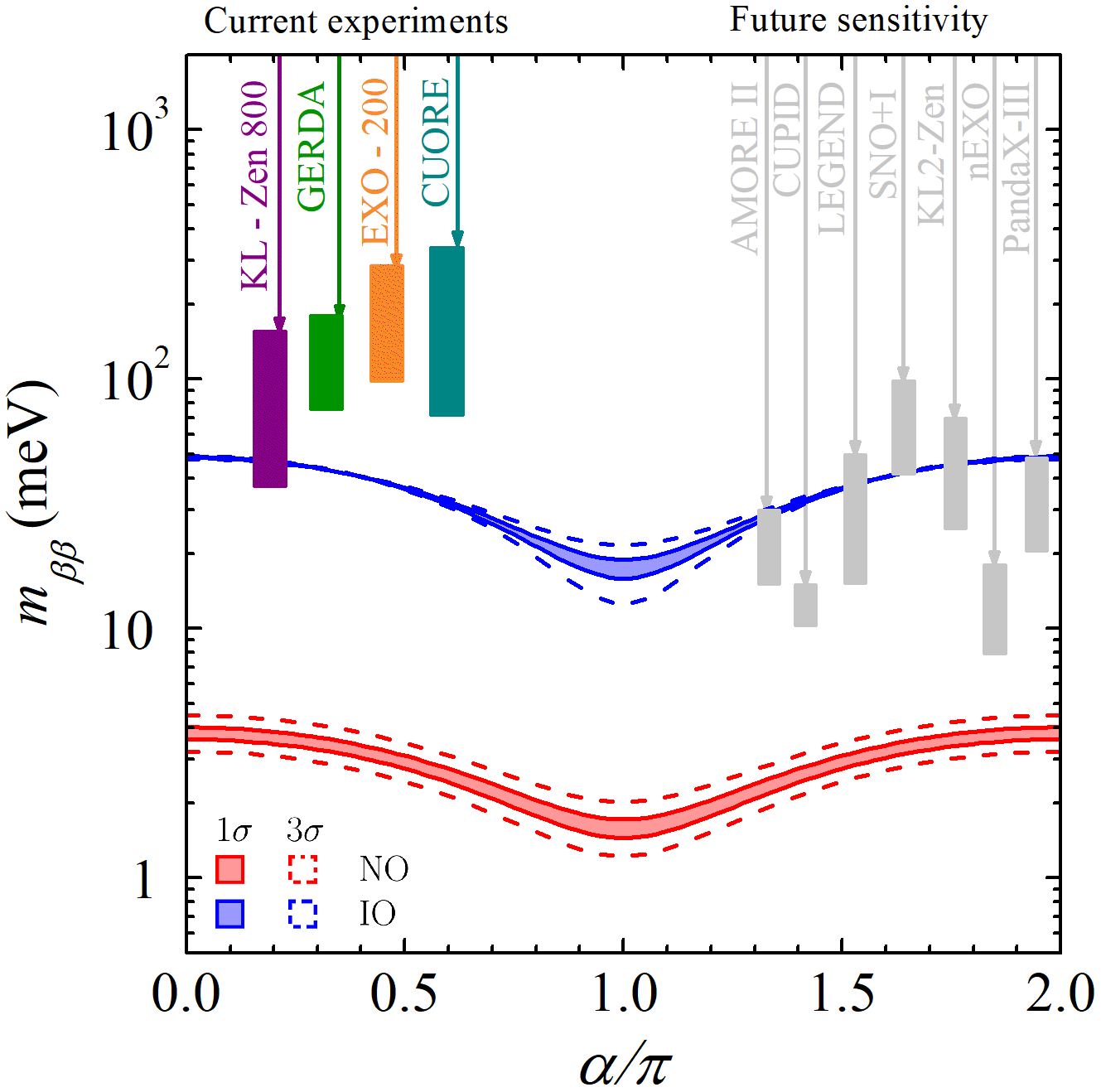}
    \caption{Allowed $m_{\beta \beta}$ regions as a function of $\alpha$ for our minimal $\nu$DFSZ models obtained using Eqs.~\eqref{eq:NOIOmbb1} and~\eqref{eq:NOIOmbb2} and varying low-energy neutrino observables within their allowed 1$\sigma$ and 3$\sigma$ intervals by neutrino oscillation data presented in Table~\ref{tab:data} (see also Ref.~\cite{deSalas:2020pgw}). Results for NO (IO) are shown in red (blue). Current most constraining experimental upper-bound ranges are represented by vertical coloured bars, for KamLAND-Zen 800~\cite{KamLAND-Zen:2022tow} (in purple), GERDA~\cite{Agostini:2020xta} (in green), EXO-200~\cite{Anton:2019wmi} (in orange) and CUORE~\cite{Adams:2019jhp} (in teal), while future sensitivities of several upcoming experiments are indicated in gray (see text for details).}
    \label{fig:mbb_alpha}
\end{figure}
As already mentioned, the pair ($\mathbf{M}_{e}$, $\mathbf{M}_\nu$), where $\mathbf{M}_\nu$ is determined from $\mathbf{M}_D$ and $\mathbf{M}_R$ using \eqref{eq:EffectiveNeutrino}, contains the same number of parameters as the number of lepton observables given by those in Table~\ref{tab:data} and the single Majorana phase $\alpha$. By construction, since there are only two RH neutrinos one of the three light neutrinos is predicted to be massless due to the missing partner nature~\cite{Schechter:1980gr} of the underlying type-I seesaw mechanism. Hence, the framework presented here is testable at future experiments looking for neutrinoless double beta decays $(\beta \beta)_{0\nu}$. The amplitude for this rare process is proportional to the effective Majorana mass parameter $m_{\beta \beta}$, which in our minimal scenario, can be written, for both mass orderings, as~\cite{Joaquim:2003pn}:
\begin{align}
\text{NO :}& \ m_{\beta \beta} = \left| \sqrt{\Delta m_{21}^2 } \ s_{12}^2 c_{13}^2 e^{-i \alpha}  + \sqrt{\left| \Delta m_{31}^2 \right|} \ s_{13}^2 \right| \; , \label{eq:NOIOmbb1} \\ 
\text{IO :}& \ m_{\beta \beta} = \left|\sqrt{\left| \Delta m_{31}^2 \right|} \ c_{12}^2  c_{13}^2 + \sqrt{\Delta m_{21}^2 +  \left| \Delta m_{31}^2 \right|} \ s_{12}^2 c_{13}^2 e^{-i \alpha}\right| \; ,
\label{eq:NOIOmbb2}
\end{align}
where it is clear that the resulting allowed regions, by the oscillation data, for $m_{\beta \beta}$ will correlate with the Majorana phase~$\alpha$, as shown in Fig.~\ref{fig:mbb_alpha}. Results for NO and IO are presented in red and blue, respectively, with the colored (dashed) regions obtained by varying the neutrino observables within their allowed 1$\sigma$ (3$\sigma$) intervals (see Table~\ref{tab:data} and Ref.~\cite{deSalas:2020pgw}). Current experimental constraints are represented by vertical colored bars, for KamLAND-Zen 800~\cite{KamLAND-Zen:2022tow} (in purple), GERDA~\cite{Agostini:2020xta} (in green), EXO-200~\cite{Anton:2019wmi} (in orange) and CUORE~\cite{Adams:2019jhp} (in teal). Also are shown in gray the future $m_{\beta\beta}$ sensitivities projected by the upcoming experiments~AMORE~II~\cite{Lee:2020rjh}, CUPID~\cite{Wang:2015raa}, LEGEND~\cite{Abgrall:2017syy}, SNO+~I~\cite{Andringa:2015tza}, KamLAND2-Zen~\cite{KamLAND-Zen:2016pfg}, nEXO~\cite{Albert:2017hjq} and PandaX-III~\cite{Chen:2016qcd}. We notice that in this minimal scenario there is no cancellation in the $(\beta \beta)_{0\nu}$ amplitude even for NO (red) neutrino masses~\cite{Barreiros:2018bju}. Namely, the effective Majorana mass parameter~\cite{Barreiros:2018bju} has a lower bound for NO neutrino masses $m_{\beta \beta} \sim 1.5$ meV for $\alpha \sim \pi$. Most interestingly for IO (blue) the predictions $m_{\beta \beta} \in [15,50]$ meV, fall within the projected sensitivities of upcoming experiments (gray bars). Overall, a positive result would not only prove the Majorana nature of neutrinos by the black-box theorem~\cite{Schechter:1981bd}, but could potentially determine the Majorana CP-violating phase~\cite{Branco:2002ie}.

Although for the sake of compatibility with fermion mass and mixing data, any lepton model can be combined with any quark model, the choice of a specific lepton model is relevant for the axion-to-photon coupling as discussed in Sec.~\ref{sec:axionphoton}.

\end{itemize}
We conclude this section by remarking that the minimal flavor models found compatible with data, eight for quarks and eleven for leptons, will result in a total of eighty-eight combined models. In what follows -- Sec.~\ref{sec:axionpheno} -- we will investigate  the properties of these models regarding axion phenomenology.

\section{Axion phenomenology}
\label{sec:axionpheno}

The growing experimental axion program encompasses a diverse array of facilities dedicated to the search for this elusive particle, most notably helioscopes and haloscopes. Alongside astrophysical and cosmological observations, the axion parameter space has been constrained by probing axion couplings to photons, electrons, and nucleons (for comprehensive reviews, see Refs.~\cite{DiLuzio:2020wdo,Adams:2022pbo}). In this section we present the axion-to-photon coupling predictions for the different models identified in the previous section, and we investigate possible constraints on axion flavor violating couplings to quarks and charged leptons~\cite{MartinCamalich:2020dfe,Alonso-Alvarez:2023wig}.

\subsection{Axion dark matter and cosmology}
\label{sec:axiondarkmatter}

The PQ solution to the strong CP problem becomes even more attractive when one considers the axion as a DM candidate, solving yet another problem the SM cannot account for. Axions are naturally light, weakly coupled with ordinary matter, nearly stable and can be non-thermally produced in the early Universe. Axion DM is conceivable in two distinct cosmological scenarios:
\begin{itemize}

    \item \textbf{Pre-inflationary:} The PQ symmetry is broken before inflation and is not subsequently restored. Axion DM is generated entirely through the misalignment mechanism~\cite{Preskill:1982cy,Abbott:1982af,Dine:1982ah}. Before inflation, different regions of the Universe have varying values for the axion field $a(\vec{x},t) = a(t) = \theta_0 f_a$, where $|\theta_0| \in [0, \pi)$, is the so-called initial misalignment angle. One of those regions is expanded to our observable Universe during inflation and the axion field is driven towards the vacuum minimum at a temperature $T \sim \Lambda_{\text{QCD}}$ around the QCD phase transition. The groundstate of the axion potential preserves CP dynamically solving the strong CP problem. The axion relic abundance generated via the misalignment mechanism is approximately given by~\cite{DiLuzio:2020wdo} 
    \begin{equation}
    \Omega_a h^2 \simeq  \Omega_\text{CDM} h^2 \frac{\theta_0^2}{2.15^2} \left(\frac{f_a}{2 \times 10^{11} \ \text{GeV}} \right)^{\frac{7}{6}} \; ,
    \label{eq:relica}
    \end{equation}
    with the observed CDM relic abundance, obtained by Planck satellite data being $\Omega_{\text{CDM}} h^2 = 0.1200 \pm 0.0012$~\cite{Planck:2018vyg}. Hence, requiring axions to account for the full CDM budget with $\theta_0 \sim \mathcal{O}(1)$ implies $f_a \sim 5 \times 10^{11} \ \text{GeV}$ or, equivalently, an axion mass around $m_a \sim 7 \ \mu\text{eV}$. As discussed in Sec.~\ref{sec:axionphoton} and shown in Fig.~\ref{fig:gagg}, this parameter space region is currently being probed by haloscope experiments. 
    
    In the pre-inflationary scenario, axions leave an imprint in primordial fluctuations, which is reflected in the cosmic microwave background~(CMB) anisotropies and large-scale structure. The resulting isocurvature fluctuations are constrained by CMB data~\cite{Beltran:2006sq}, leading to an upper bound on the inflationary scale~\cite{DiLuzio:2020wdo}. In this work we do not provide a concrete inflationary mechanism where this CMB constraint could be applied. The study of inflation in our minimal flavor $\nu$DFSZ models deserves future investigation, along the lines of e.g. Refs.~\cite{Sopov:2022bog,RVolkas:2023jiv,Matlis:2023eli}.

    \item \textbf{Post-inflationary:} The PQ symmetry is broken after inflation, leading to an observable Universe divided into patches with different values of the axion field. The value of the initial misalignment angle $\theta_0$ is no longer free and, through statistical average, the up-to-date value $\left<\theta_0^2\right> \simeq 2.15^2$~\cite{DiLuzio:2020wdo} has been obtained. Hence, compared to the pre-inflationary scenario, this case is, in principle, more predictive. The requirement  $\Omega_a h^2 = \Omega_{\text{CDM}} h^2 = 0.12$, leads to a prediction for the axion scale $f_a$ (or equivalently mass $m_a$) [see Eq.~\eqref{eq:relica}]. In fact, if only the misalignment mechanism is at the origin of axion abundance, the upper bound of $f_a \lesssim 2 \times 10^{11}$ GeV guarantees that DM is not overproduced. However, the picture is much more complicated since topological defects, strings and DWs, will also contribute to the total axion relic abundance $\Omega_a h^2$~\cite{Bennett:1987vf,Levkov:2018kau,Gorghetto:2018myk,Buschmann:2019icd}. Namely, the complex non-linear dynamics of the resulting network of axion strings and DWs must be carefully resolved through numerical simulations, with their precise contribution remaining an open question and an active area of ongoing research. A detailed analysis of this lies beyond the scope of the present work.

    The cosmological DW problem is absent if $N_{\text{DW}} = 1$, where $N_{\text{DW}}$ corresponds to the vacuum degeneracy stemming from the residual $\mathcal{Z}_{N}$ symmetry, left unbroken from non-perturbative QCD-instanton effects that anomalously break U(1)$_{\text{PQ}}$. This occurs for the quark models Q$_{2,4}$ of Table~\ref{tab:quarkcharges} for any lepton model in Table~\ref{tab:leptoncharges}, as long as $V_{\rm PQ}$ in Eq.~\eqref{eq:VPQoperator} is the one with $\chi_\sigma = 1$ (cubic model). All remaining cases lead to DWs in the early Universe~\cite{Lazarides:2018aev}. 

    Although there are no DWs for $N_{\text{DW}}=1$, axionic string networks can still form. Numerical simulations predict the axion scale $f_a$ to be in the range $[5 \times 10^9,3 \times 10^{11}]$~GeV, in order for the axion to account for the observed CDM abundance, i.e. $\Omega_a h^2 = \Omega_{\text{CDM}} h^2$~\cite{Kawasaki:2014sqa,Klaer:2017ond,Gorghetto:2020qws,Buschmann:2021sdq,Benabou:2024msj}~\footnote{Here we adopt a deliberately conservative range, $f_a \in [5\times 10^9,\,3\times 10^{11}] \,\mathrm{GeV}$ (i.e. $m_a \in [19,1140]\,\mu\mathrm{eV}$), which covers the range of values quoted in Refs.~\cite{Kawasaki:2014sqa,Klaer:2017ond,Gorghetto:2020qws,Buschmann:2021sdq,Benabou:2024msj}. Recent analyses, e.g. Ref.~\cite{Benabou:2024msj}, obtain a narrower window, $m_a \in (45,65)\,\mu\mathrm{eV}$, lying entirely within our interval.}. Note that, in Ref.~\cite{Cox:2023squ}, all quark flavor patterns with $N_{\text{DW}}=1$ were identified in the DFSZ model. Our partial results for the quark sector agree with the ones found in that reference.

\end{itemize}
%

\subsection{Axion-to-photon coupling}
\label{sec:axionphoton}

%
\begin{table}[t!]
\renewcommand*{\arraystretch}{1.4}
\begin{tabular}{|c|c|}
\hline
Models & $E/N$ \\ \hline
$(\text{Q}_\text{2}^\text{II},\text{L}_\text{1-9})$ & -10/3  \\
$(\text{Q}_\text{2}^\text{II},\text{L}_\text{10,11})$ & -4/3  \\
$(\text{Q}_\text{1}^\text{II},\text{L}_\text{1-9})$ & -1/3  \\
$(\text{Q}_\text{1,2/3,4}^\text{II,I/II,I},\text{L}_\text{10,11/1-9})$ & 2/3  \\
$(\text{Q}_\text{3}^\text{II},\text{L}_\text{10,11})$ & 5/3  \\
$(\text{Q}_\text{1,2/3,4}^\text{I},\text{L}_\text{1-9/10,11})$ & 8/3 \\
$(\text{Q}_\text{3}^\text{I},\text{L}_\text{1-9})$ & 11/3 \\
$(\text{Q}_\text{4}^\text{II},\text{L}_\text{10,11})$ & 14/3 \\
$(\text{Q}_\text{4}^\text{II},\text{L}_\text{1-9})$ & 20/3 \\
\hline
\end{tabular}
\caption{$E/N$ values for the quark and lepton model combinations identified in Sec.~\ref{sec:symmetries} (see Tables~\ref{tab:quarkcharges} and~\ref{tab:leptoncharges}).}
\label{tab:EN}
\end{table}
We now examine how to probe the various scenarios of Tables~\ref{tab:quarkcharges} and~\ref{tab:leptoncharges} through their corresponding axion-to-photon couplings $g_{a \gamma \gamma}$. Using NLO chiral Lagrangian techniques, the following result has been obtained for $g_{a \gamma \gamma}$~\cite{GrillidiCortona:2015jxo} 
\begin{align}
g_{a \gamma \gamma} &= \frac{\alpha_e}{2 \pi f_a} \left[\frac{E}{N} - 1.92(4) \right] \; .
\label{eq:gagg}
\end{align}
The ratio $E/N$ is the model-dependent contribution, with $E$ being the electromagnetic anomaly factor, given by 
\begin{equation}
E = 2 \sum_f \left(\chi_f^L - \chi_f^R \right) q^2_f \; ,
\label{eq:E}
\end{equation}
where $q_f$ is the electric charge of the fermion $f$ transforming under U(1)$_{\text{PQ}}$ with charges $\chi_f^{L,R}$. To compute $E$, both quark and lepton PQ charges need to be specified. The eighty eight combinations of Q and L models (see Tables~\ref{tab:quarkcharges} and~\ref{tab:leptoncharges}) lead to nine distinct $E/N$ values, as shown in Table~\ref{tab:EN} for the minimal flavored PQ $\nu$DFSZ models identified in Sec.~\ref{sec:symmetries}. The combinations $(\text{Q}_\text{1,2/3,4}^\text{I},\text{L}_\text{1-9/10,11})$ and $(\text{Q}_\text{1,2/3,4}^\text{II,I/II,I},\text{L}_\text{10,11/1-9})$ lead to the same $E/N$ as in the DFSZ-I and DFSZ-II schemes~\cite{Zhitnitsky:1980tq,Dine:1981rt}, namely $E/N=8/3$ and $E/N=2/3$, respectively.

\begin{figure}[!t]
    \centering
      \includegraphics[scale=0.55]{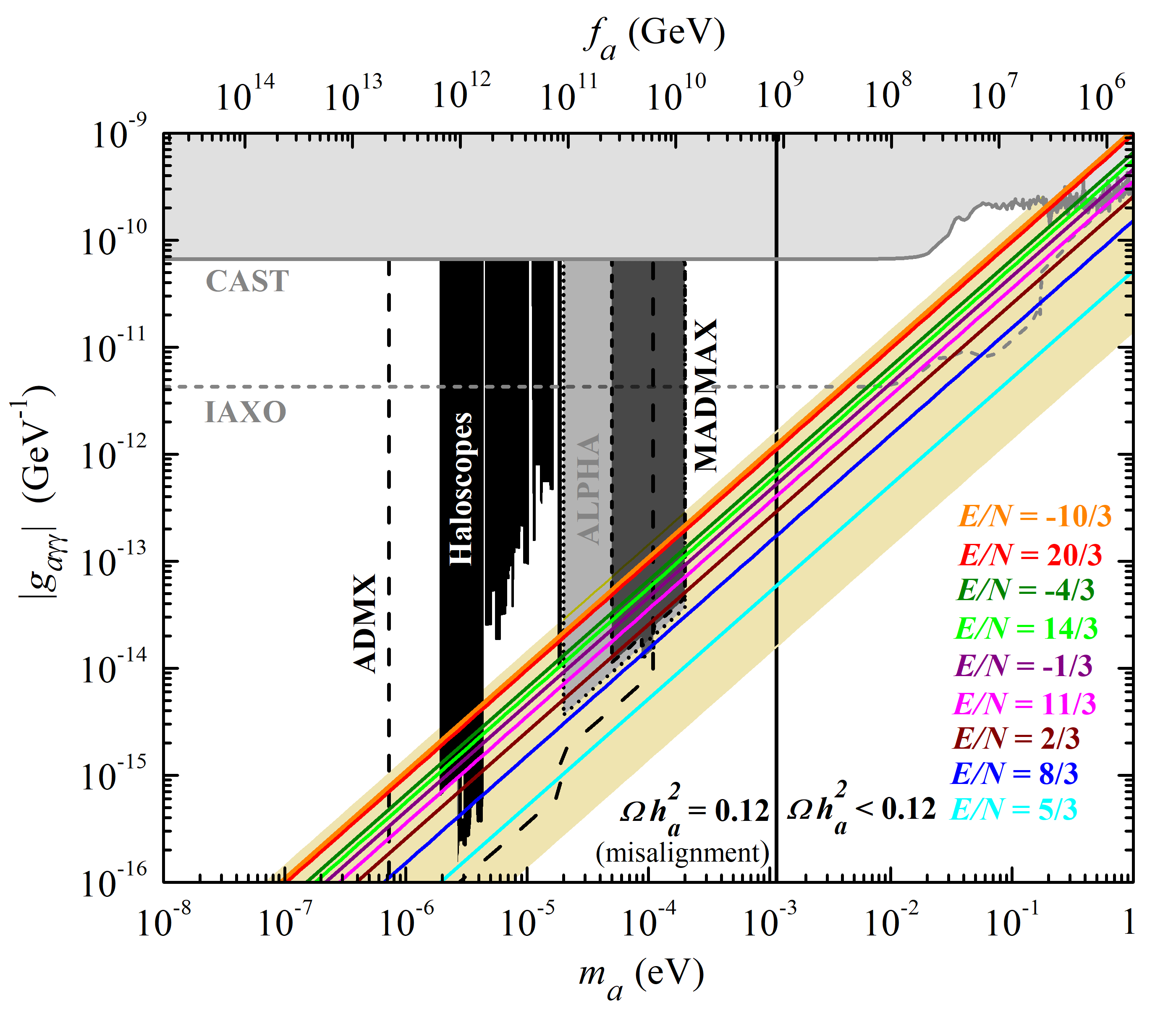}
    \caption{Axion-to-photon coupling $|g_{a \gamma \gamma}|$ versus axion mass $m_a$ (bottom axis) and decay constant $f_a$ (top axis). Colored solid lines indicate $E/N$ values for the different models identified in Sec.~\ref{sec:symmetries} (see Table~\ref{tab:EN}). The yellow shaded region refers to the usual QCD axion window~\cite{DiLuzio:2016sbl}. Current constraints from helioscopes like CAST~\cite{CAST:2017uph} exclude the light-gray shaded region, while haloscopes like ADMX~\cite{ADMX:2018gho,ADMX:2019uok,ADMX:2021nhd}, RBF~\cite{DePanfilis:1987dk}, CAPP~\cite{CAPP:2020utb} and HAYSTAC~\cite{HAYSTAC:2020kwv}, rule out the black region. Projected sensitivities of IAXO~\cite{Shilon_2013}, ADMX~\cite{Stern:2016bbw}, MADMAX~\cite{Beurthey:2020yuq} and ALPHA~\cite{ALPHA:2022rxj} are indicated by the dashed gray, dashed black, short-dash black (dark-gray shaded region) and dotted (light-gray shaded region) contours, respectively. To the right of the black vertical line, axion DM is under-abundant. In the left region $\Omega h^2_a = 0.12$, for the pre-inflationary case featuring the misalignment mechanism.}
    \label{fig:gagg}
\end{figure}
In Fig.~\ref{fig:gagg}, we display $|g_{a\gamma \gamma}|$ in terms of the axion mass~$m_a$ (bottom axis) and decay constant~$f_a$ (top axis), showing current bounds and future sensitivities from helioscopes and haloscopes. The colored oblique lines indicate the $|g_{a\gamma \gamma}|$ predictions for the model combinations of Table~\ref{tab:EN}. A few comments are in order:
\begin{itemize}
    \item The maximum and minimum values for $|g_{a\gamma \gamma}|$ correspond to the $(\text{Q}_\text{2}^\text{II},\text{L}_\text{1-9})$ and $(\text{Q}_\text{3}^\text{II},\text{L}_\text{10,11})$ models, leading to $E/N=-10/3$ (orange line) and $E/N=5/3$ (cyan line), respectively. Naturally, the distinct $|g_{a\gamma \gamma}|$ predictions are within the yellow shaded region corresponding to the usual QCD axion window $|E/N -1.92| \in [0.07,7]$~\cite{DiLuzio:2016sbl} -- see Eq.~\eqref{eq:E}.

    \item Helioscopes aim at detecting the solar axion flux, with the CERN Solar Axion Experiment~(CAST)~\cite{CAST:2017uph} currently excluding $|g_{a \gamma \gamma}| \geq 0.66 \times 10^{-10} \ \text{GeV}^{-1}$ for masses $m_a \leq 20$ meV (light-gray-shaded region). For $m_a \gtrsim 0.2$~eV we can roughly say that the only models not excluded by CAST are those with $E/N=2/3,\,8/3$ and 5/3. The upcoming International Axion Observatory~(IAXO)~\cite{Shilon_2013} is expected to probe the axion-to-photon coupling $g_{a \gamma \gamma}$ down to $(10^{-12}-10^{-11}) \ \text{GeV}^{-1}$ for $m_a \sim 0.1$ eV (gray-dashed contour). This will scrutinize all our models except for $(\text{Q}_\text{3}^\text{II},\text{L}_\text{10,11})$ that leads to the minimal $|g_{a\gamma \gamma}|$ value for $E/N=5/3$ (cyan line).

    \item Haloscopes aim at detecting non-relativistic axions under the assumption that they make up all the CDM observed in the Universe (see Sec.~\ref{sec:axiondarkmatter}). These experiments include ADMX~\cite{ADMX:2018gho,ADMX:2019uok,ADMX:2021nhd}, RBF~\cite{DePanfilis:1987dk}, CAPP~\cite{CAPP:2020utb} and HAYSTAC~\cite{HAYSTAC:2020kwv}, which exclude the black shaded region. As discussed in Sec.~\ref{sec:axiondarkmatter}, the QCD axion can account for the observed DM relic abundance via the misalignment mechanism in the pre-inflationary scenario, in the region to the left of the black vertical line [see Eq.~\eqref{eq:relica}]. In contrast, the right region leads to under-abundant axion DM. Out of all the haloscope experiments, the most impressive and promising is the Axion Dark Matter eXperiment~(ADMX). Namely, ADMX currently excludes a considerable part of the parameter space of our scenarios, for axion masses $m_a \sim 3 \ \mu$eV, corresponding to $f_a \sim 10^{12}$ GeV and $g_{a \gamma \gamma}$ down to $(2 \times 10^{-16}) \ \text{GeV}^{-1}$. Again, model $(\text{Q}_\text{3}^\text{II},\text{L}_\text{10,11})$ (cyan line) is an exception. Future ADMX generations (dashed black contour) are expected to probe all our models for masses $1 \ \mu \text{eV} \lesssim m_a \lesssim 20 \ \mu$eV or equivalently scales $5 \times 10^{11} \ \text{GeV} \lesssim f_a \lesssim 10^{13}$ GeV, with $g_{a \gamma \gamma}$ down to $(10^{-16}-10^{-15}) \ \text{GeV}^{-1}$. ALPHA~\cite{Lawson:2019brd,Wooten:2022vpj,ALPHA:2022rxj} and MADMAX~\cite{Beurthey:2020yuq} are projected to probe the mass region $[20,200] \ \mu \text{eV}$ (dash contour and dark-gray region) and $[50,200] \ \mu \text{eV}$ (dotted contour and light-gray region), respectively, covering all our frameworks except for $(\text{Q}_\text{1,2/3,4}^\text{I},\text{L}_\text{1-9/10,11})$ (blue line) and $(\text{Q}_\text{3}^\text{II},\text{L}_\text{10,11})$ (cyan line), that lead to $E/N=8/3$ and $E/N=5/3$, respectively~(see Table~\ref{tab:EN}).
    
\end{itemize}
Overall, haloscope and helioscope experiments provide an important way to probe our models, complementary to other new physics searches in the flavor sector, as it will be explored in the upcoming section.

\subsection{Flavor-violating axion couplings to fermions}
\label{sec:axionflavorviolating}

The strong hierarchy among the singlet and doublet VEVs $v_\sigma \gg v_1,v_2$, leads to a decoupling between the scalar particles of the softly-broken U(1) 2HDM and the singlet~\cite{Sopov:2022bog}. In this framework, the presence of new charged and neutral scalars that couple with fermions introduces new physics contributions in flavor processes. The tree-level FCNCs will be controlled by the matrices,
\begin{align}
    \mathbf{N}_{f} &= 
    \frac{v}{\sqrt{2}} 
    \mathbf{U}_L^{f \dagger}
    \left(s_\beta\Y_1^{f} - c_\beta\Y_2^{f} \right)
    \mathbf{U}_R^{f} \; ,
\label{eq:NMatrices}
\end{align}
with the Yukawa couplings $\Y_{1,2}^{f}$ ($f=u,d,e$) and unitary rotations $\mathbf{U}_{L,R}^{f}$ defined as in Eqs.~\eqref{eq:Lyuk2hdm}, \eqref{eq:massdiag} and~\eqref{eq:leptonOrdering}. As shown in Refs.~\cite{Correia:2019vbn,Camara:2020efq,Rocha:2024twm}, the maximally restrictive U(1) flavor symmetries -- promoted to a PQ symmetry in this work -- can, in some cases, naturally suppress such dangerous FCNCs effects. In fact, the mass matrices labelled ``$5$" in Table~\ref{tab:Matrices}, feature an isolated non-zero entry in a specific row and column, corresponding to the mass of a particular fermion—referred to as the ``decoupled state". Therefore, for each model with a type ``$5$" matrix, we can have three decoupled states, which we identify with a superscript on the model tag. For instance, if $u$ is the decoupled state in $\text{Q}_1$, we refer to it as $\text{Q}_1^u$~\footnote{Since permuting the Yukawa couplings results in the same texture for $\mathbf{N}_{f}$, we omit the I/II superscript in this section for simplicity.}. Note that models with decoupled states have zero entries in the $\mathbf{N}_{f}$ matrices of Eq.~\eqref{eq:NMatrices}, which control the strength of FCNCs,
\begin{align}
    \text{Q}^{d}_{1-2},\text{Q}^{u}_{3-4},\text{L}^e_{10-11}&: \mathbf{N}_{d,u,e} \sim \begin{pmatrix}
    \times & 0 & 0 \\
    0 & \times & \times \\
    0 & \times & \times \\
    \end{pmatrix} \; , \; \nonumber
    \\
    \text{Q}^{s}_{1-2},\text{Q}^{c}_{3-4},\text{L}^\mu_{10-11} &: \mathbf{N}_{d,u,e} \sim \begin{pmatrix}
    \times & 0 & \times  \\
    0 & \times & 0 \\
    \times & 0 & \times \\
    \end{pmatrix} \; , \;  
    \text{Q}^{b}_{1-2},\text{Q}^{t}_{3-4},\text{L}^\tau_{10-11} : \mathbf{N}_{d,u,e} \sim \begin{pmatrix}
    \times & \times & 0  \\
    \times & \times & 0 \\
    0 & 0 & \times \\
    \end{pmatrix} \; .
\label{eq:Nudtextures}
\end{align}
Thanks to this symmetry-induced FCNCs suppression mechanism, it was found in our previous work~\cite{Rocha:2024twm}, that some of our models comply with stringent quark sector flavor constraints, with new scalars below the TeV scale, within the reach of current experiments such as the LHC and testable at future facilities. Specifically, for down/strange decoupled models strong constraints stemming from the mass differences of the meson-antimeson systems $K$, $B_d$ and $B_s$, as well as CP violation encoded in the $\epsilon_K$ parameter, are automatically fulfilled, leading to neutral BSM scalar masses as low as $300$ GeV.

\begin{table}[t!]
    \renewcommand*{\arraystretch}{1.5}
    \centering
    \begin{tabular}{|c|c|c|c|c|}
    \hline 
     Most Restrictive Processes & $\; \alpha, \beta \;$   & $\mathbf{F}^V_{\alpha \beta}$ (GeV)  & $\mathbf{F}^A_{\alpha \beta}$ (GeV)  & Experiment \& Ref. \\
    \hline
    $K^+\rightarrow \pi^+ a$ & $\; s,d \;$ & \renewcommand{\arraystretch}{1.0}\begin{tabular}[c]{@{}c@{}} $6.8\times10^{11}$ \\ $(2\times10^{12})$\end{tabular} & -  &  \renewcommand{\arraystretch}{1.0}\begin{tabular}[c]{@{}c@{}} E949 \& E787~\cite{E949:2007xyy} \\ (NA62 \& KOTO~\cite{MartinCamalich:2020dfe})\end{tabular}
     \\
    \hline
    $\Lambda\rightarrow$ $n$ $a$ (Supernova)  & $\; s,d \;$ & $7.4\times10^9$ & $5.4\times 10^9$ & PDG~\cite{ParticleDataGroup:2018ovx}
     \\
    \hline
    $D^+\rightarrow \pi^+ a$   & $\; c,u \;$ & \renewcommand{\arraystretch}{1.0}\begin{tabular}[c]{@{}c@{}} $9.7\times10^{7}$ \\ $(5\times10^{8})$\end{tabular} & - &     \renewcommand{\arraystretch}{1.0}\begin{tabular}[c]{@{}c@{}} CLEO~\cite{CLEO:2008ffk}  \\ (BES III~\cite{MartinCamalich:2020dfe}) \end{tabular}
     \\
    \hline
    $\Lambda_c \rightarrow$ $p$ $a$ & $\; c,u \;$ & \renewcommand{\arraystretch}{1.0}\begin{tabular}[c]{@{}c@{}} $1.4\times10^{5}$ \\ $(2\times10^{7})$\end{tabular} & \renewcommand{\arraystretch}{1.0}\begin{tabular}[c]{@{}c@{}} $1.2\times10^{5}$ \\ $(2\times10^{7})$\end{tabular} &  
    \renewcommand{\arraystretch}{1.0}\begin{tabular}[c]{@{}c@{}} PDG~\cite{ParticleDataGroup:2018ovx} \\ (BES III~\cite{MartinCamalich:2020dfe}) \end{tabular}
    \\
    \hline
    $B^{+,0}\rightarrow K^{ +,0}a$  & $\; b,s \;$ & \renewcommand{\arraystretch}{1.0}\begin{tabular}[c]{@{}c@{}} $3.3\times10^{8}$ \\ $(3\times10^{9})$\end{tabular} & - &     \renewcommand{\arraystretch}{1.0}\begin{tabular}[c]{@{}c@{}} BABAR~\cite{BaBar:2013npw}   \\ (BELLE II~\cite{MartinCamalich:2020dfe}) \end{tabular}
    \\
    \hline
    $B^{+,0}\rightarrow K^{\ast +,0}a$ & $\; b,s \;$ & - & \renewcommand{\arraystretch}{1.0}\begin{tabular}[c]{@{}c@{}} $1.3\times10^{8}$ \\ $(1\times10^{9})$\end{tabular} &   \renewcommand{\arraystretch}{1.0}\begin{tabular}[c]{@{}c@{}} BABAR~\cite{BaBar:2013npw}   \\ (BELLE II~\cite{MartinCamalich:2020dfe}) \end{tabular}
    \\
    \hline
    $B^{+}\rightarrow \pi^{+}a$ & $\; b,d \;$ & \renewcommand{\arraystretch}{1.0}\begin{tabular}[c]{@{}c@{}} $1.1\times10^{8}$ \\ $(3\times10^{9})$\end{tabular} & - &  \renewcommand{\arraystretch}{1.0}\begin{tabular}[c]{@{}c@{}} BABAR~\cite{BaBar:2004xlo}   \\ (BELLE II~\cite{MartinCamalich:2020dfe}) \end{tabular} 
    \\
    \hline
    $B^{+,0}\rightarrow \rho^{ +,0}a$ & $\; b,d \;$ & - &  $(1\times10^{9})$ & (BELLE II~\cite{MartinCamalich:2020dfe})
    \\
    \hline
    $K^{+}\rightarrow \pi^{ +}a$ (loop) & $\; t,u \;$ & $3\times10^8$ &  $3\times10^8$ &  E949 \& E787~\cite{E949:2007xyy}
    \\
    \hline
    $K^{+}\rightarrow \pi^{ +}a$ (loop) & $\; t,c \;$ & $7\times10^8$ &  $7\times10^8$ &  E949 \& E787~\cite{E949:2007xyy}
    \\
    \hline
\end{tabular}
\caption{The most relevant $90\%$ CL lower bounds on the scales of flavor-violating axion-quark couplings $\mathbf{F}_{\alpha \beta}^{V,A}$ defined in Eq.~\eqref{eq:Fdef}, with future projections in parentheses (obtained from Ref.~\cite{MartinCamalich:2020dfe}). Constraints labeled as “loop" originate from radiative corrections contributing to the relevant flavor process.
}
\label{tab:QuarkConstraints}
\end{table}
\begin{table}[t!]
    \renewcommand*{\arraystretch}{1.5}
    \centering
    \begin{tabular}{|c|c|c|c|c|}
    \hline 
     Most Restrictive Processes & $\; \alpha,\beta \;$   & $\mathbf{F}_{\alpha\beta}$ (GeV)    & Experiment \& Ref. \\
    \hline
    Star Cooling & $\; e,e \;$ & $4.6\times10^9$  & WDs~\cite{MillerBertolami:2014rka} 
\\
    \hline
    Red Giants & $\; e,e \;$ & $6.4\times10^9$  & TRGB~\cite{Capozzi:2020cbu,Bottaro:2023gep}
     \\
    \hline
   Star Cooling  & $\; \mu,\mu \;$  & $3.2\times 10^7$ & SN$1987$a$_{\mu \mu}$~\cite{Bollig:2020xdr,Croon:2020lrf,Caputo:2021rux}
     \\
    \hline
    $\mu \rightarrow$ $e$  $a$ $\gamma$   & $\; \mu,e \;$ & $(5.1 -  8.3)\times10^8$  & Crystal Box~\cite{Bolton:1988af}
     \\
    \hline
    $\tau \rightarrow$ $e$  $a$  & $\; \tau,e \;$ & \renewcommand{\arraystretch}{1.0}\begin{tabular}[c]{@{}c@{}} $4.3\times10^{6}$ \\ $(7.7\times10^{7})$\end{tabular}  &  \renewcommand{\arraystretch}{1.0}\begin{tabular}[c]{@{}c@{}} ARGUS~\cite{ARGUS:1995bjh} \\ (BELLE II~\cite{Calibbi:2020jvd}) \end{tabular}
    
    \\
    \hline
    $\tau \rightarrow$ $\mu$  $a$   & $\; \tau,\mu \;$ & \renewcommand{\arraystretch}{1.0}\begin{tabular}[c]{@{}c@{}} $3.3\times10^{6}$ \\ $(4.9\times10^{7})$\end{tabular}  & \renewcommand{\arraystretch}{1.0}\begin{tabular}[c]{@{}c@{}} ARGUS~\cite{ARGUS:1995bjh} \\ (BELLE II~\cite{Calibbi:2020jvd})  \end{tabular}
    \\
    \hline
\end{tabular}
\caption{The most relevant $95\%$ CL, and $90\%$ CL for the Crystal Box experiment, lower bounds on the scales of flavor-violating axion-lepton couplings $\mathbf{F}_{\alpha \beta}$ defined in Eq.~\eqref{eq:Fdef}, with future projections in parentheses (obtained from Ref.~\cite{Calibbi:2020jvd}).}
\label{tab:LeptonConstraints}
\end{table}

In the $\nu$DFSZ model, the axion also has flavor-violating couplings to fermions. Namely, writing  Eq.~\eqref{eq:Lyuk2hdm} in the mass basis~\eqref{Eq:MassMatrices}, modifies the fermion kinetic terms resulting in the effective axion-fermion interactions~Diagonal vector couplings are unphysical and are therefore set to zero~\cite{MartinCamalich:2020dfe,Calibbi:2020jvd}.,
\begin{align}
    \mathcal{L}_{aff} = \frac{\partial_\mu a}{2 f_a} \overline{f_\alpha} \gamma^\mu \left(\mathbf{C}^{V, f}_{\alpha \beta} + \mathbf{C}^{A, f}_{\alpha \beta} \gamma_5 \right) f_\beta \; 
    , 
\label{eq:axionFermionLagrangian}
\end{align}
where
\begin{align}
    \mathbf{C}^{V,f} = \frac{1}{N} \left(\mathbf{U}^{f \dagger}_R \boldsymbol{\chi}^{R}_f \mathbf{U}_R^f + \mathbf{U}^{f \dagger}_L \boldsymbol{\chi}^{L}_f \mathbf{U}_L^f \right) \; , \;  \mathbf{C}^{A,f} = \frac{1}{N} \left(\mathbf{U}^{f \dagger}_R \boldsymbol{\chi}^{R}_f \mathbf{U}_R^f -\mathbf{U}^{f \dagger}_L \boldsymbol{\chi}^{L}_f \mathbf{U}_L^f \right) \; ,
\label{eq:axionFermionCoupling}
\end{align}
with $\boldsymbol{\chi}^{L,R}_f = \diag(\chi_1,\chi_2,\chi_3)^{L,R}_f$. In Tables~\ref{tab:QuarkConstraints} and~\ref{tab:LeptonConstraints}, we provided the most restrictive constraints on
\begin{align}
\mathbf{F}_{\alpha \beta}^{V,A} \equiv 2f_a/|\mathbf{C}_{\alpha \beta}^{V,A}| \;, \; \mathbf{F}_{\alpha \beta} \equiv 2f_a/\sqrt{|\mathbf{C}_{\alpha \beta}^{V}|^2 + |\mathbf{C}_{\alpha \beta}^{A}|^2} \; ,
\label{eq:Fdef}
\end{align}
for quarks~\cite{MartinCamalich:2020dfe,Alonso-Alvarez:2023wig} and leptons~\cite{Calibbi:2020jvd}, respectively (for a recent review on flavor-violating constraints see Ref.~\cite{MartinCamalich:2025srw}). In complete analogy with Higgs FCNCs, flavored-PQ symmetries will also control axion flavor-violating couplings. In fact, for 5-type mass matrix structure, the struture of $\mathbf{C}^{V,A}$ mirrors that of the $\mathbf{N}_f$ matrices [see Eq.~\eqref{eq:Nudtextures}], i.e.
\begin{align}
    \text{Q}^{d}_{1-2},\text{Q}^{u}_{3-4},\text{L}^e_{10-11}&: \mathbf{C}^{d,u,e} \sim \begin{pmatrix}
    \times & 0 & 0 \\
    0 & \times & \times \\
    0 & \times & \times \\
    \end{pmatrix} \; , \; \nonumber
    \\
    \text{Q}^{s}_{1-2},\text{Q}^{c}_{3-4},\text{L}^\mu_{10-11} &: \mathbf{C}^{d,u,e} \sim \begin{pmatrix}
    \times & 0 & \times  \\
    0 & \times & 0 \\
    \times & 0 & \times \\
    \end{pmatrix} \; , \;  
    \text{Q}^{b}_{1-2},\text{Q}^{t}_{3-4},\text{L}^\tau_{10-11} : \mathbf{C}^{d,u,e} \sim \begin{pmatrix}
    \times & \times & 0  \\
    \times & \times & 0 \\
    0 & 0 & \times \\
    \end{pmatrix} \; ,
\label{eq:Nudtextures}
\end{align}
where for simplicity we omitted the $V,A$ superscript. Thanks to this mechanism, some flavor-violating constraints are automatically satisfied, i.e. they do not bound the axion mass~\cite{Cox:2023squ}. Specifically, a model with either $\alpha$ or $\beta$ decoupled is consistent with the bounds set on $|\mathbf{C}_{\alpha \neq \beta}^{V,A}|$ (see Tables~\ref{tab:QuarkConstraints} and~\ref{tab:LeptonConstraints}), for any axion mass. However, note that this mechanism does not apply to flavor-conserving constraints, i.e. for $\alpha = \beta$. On the other hand, the mass matrices labelled ``$4$'' lead to $\mathbf{C}^{V,A}$ matrices without zero entries. All axion models listed in Tables~\ref{tab:quarkcharges} and~\ref{tab:leptoncharges}, share some interesting properties, specifically:
\begin{itemize}
    
    \item The off-diagonal entries of $\mathbf{C}^{V,f}$ and $\mathbf{C}^{A,f}$, defined in Eq.~\eqref{eq:PQsymmetry}, are independent of the angle $\beta$. This is evident since the PQ charges of  different $f$-fermion families depend equally on this parameter, implying that only the diagonal couplings exhibit $\beta$ dependence. This results from unitarity of the mixing matrices in Eqs.~\eqref{eq:massdiag} and~\eqref{eq:leptonOrdering}.

    \item In our models, it can be shown that all the axion couplings to fermions can be made real. Thus, certain strong constraints do not apply, as those coming from the $D-\bar{D}$ mixing CP-violating phase~\cite{MartinCamalich:2020dfe}. Instead, the weaker CP-conserving constraint applies.

    \item Yukawa permutations in the quark models do not affect the physical axion-fermion couplings. Therefore, in this section, models that differ only by a Yukawa permutation, such as $\text{Q}_1^\text{I}$ and $\text{Q}_1^\text{II}$, can be treated together (we denote them as $\text{Q}_{1}$).

    \item Lepton models allowing for the cubic ($\chi_\sigma=1$) and quartic ($\chi_\sigma=1/2$) $V_{\rm PQ}$ yield the same $\mathbf{C}^{V,A}$ couplings. Consequently, in this section, lepton models with the same textures can be studied simultaneously, independently from the form of $V_{\rm PQ}$.
    
\end{itemize}
\begin{figure}[!t]
    \centering
    \includegraphics[scale=0.55]{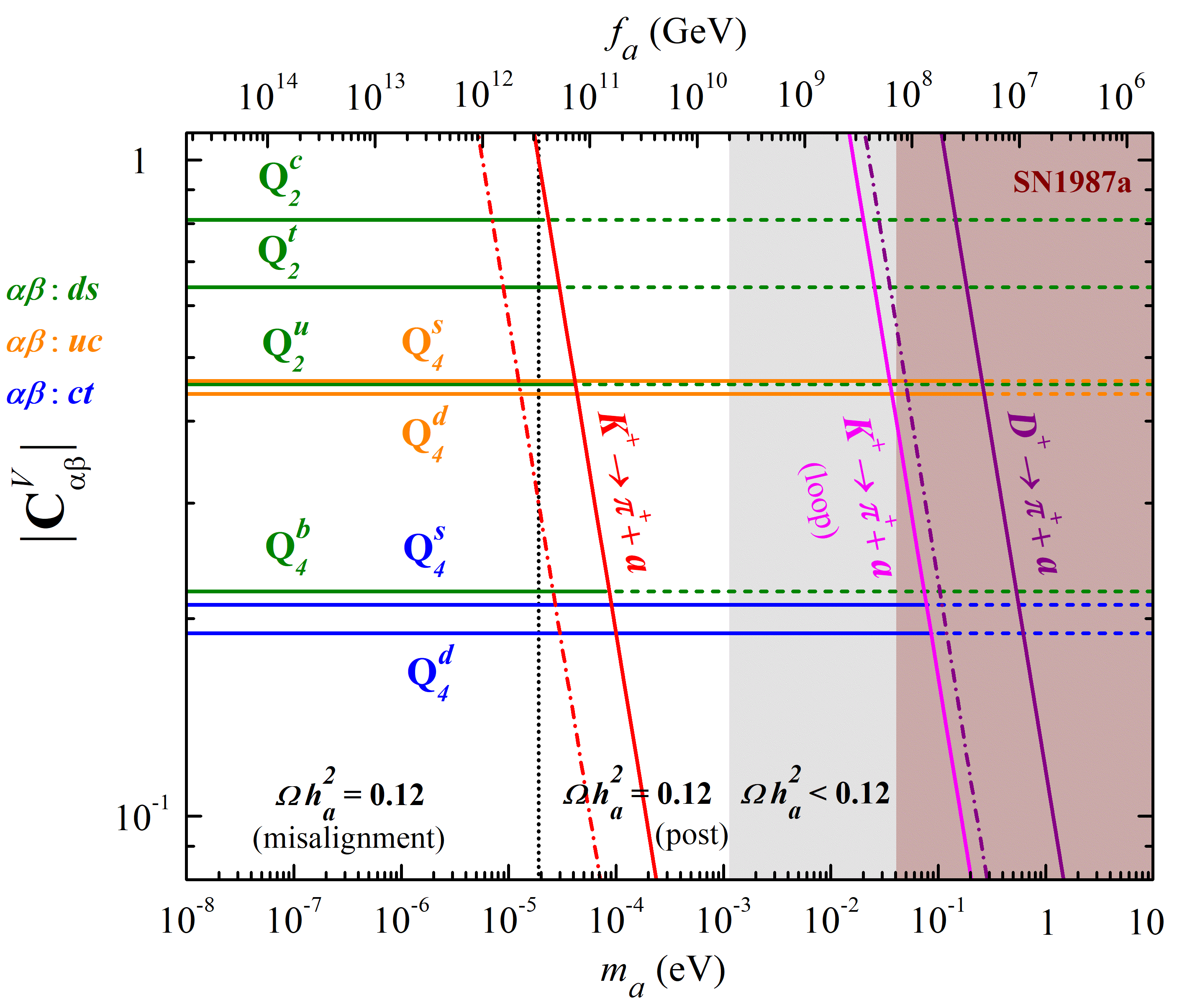}
    \caption{Vector flavor-violating axion-quark couplings $|\mathbf{C}^V_{\alpha\beta}|$, versus $m_a$ (bottom axis) and $f_a$ (top axis). We present the most restricted couplings $\alpha \beta = ds$ ($uc$) [$ct$], indicated by horizontal green (orange) [blue] lines, for models featuring $N_{\text{DW}} =1$ (see Table~\ref{tab:quarkcharges} and Sec.~\ref{sec:axiondarkmatter}). The dashed part of the lines are currently excluded by the red (purple) [magenta] oblique bound from $K^+ \rightarrow \pi^+ + a$ ($D^+ \rightarrow \pi^+ + a$) [$K^+ \rightarrow \pi^+ + a$ (loop)]~\cite{MartinCamalich:2020dfe,Alonso-Alvarez:2023wig} (see Table~\ref{tab:QuarkConstraints} and text for details). Future experimental sensitivities are indicated by oblique dash-dotted lines. SN1987a excludes the purple shaded region~\cite{Carenza:2019pxu}. In the gray shaded region axion DM is underabundant, while in the white region $\Omega h^2_a = 0.12$, for the pre-inflationary case. Considering the post-inflationary scenario for $N_{\text{DW}} =1$ the axion mass is restricted to the white region to the right of the vertical dotted black line where $\Omega h^2_a = 0.12$~\cite{Buschmann:2021sdq,Gorghetto:2020qws,Klaer:2017ond,Kawasaki:2014sqa}.}
    \label{fig:quarks}
\end{figure}
\begin{figure}[!t]
    \centering
    \includegraphics[scale=0.55]{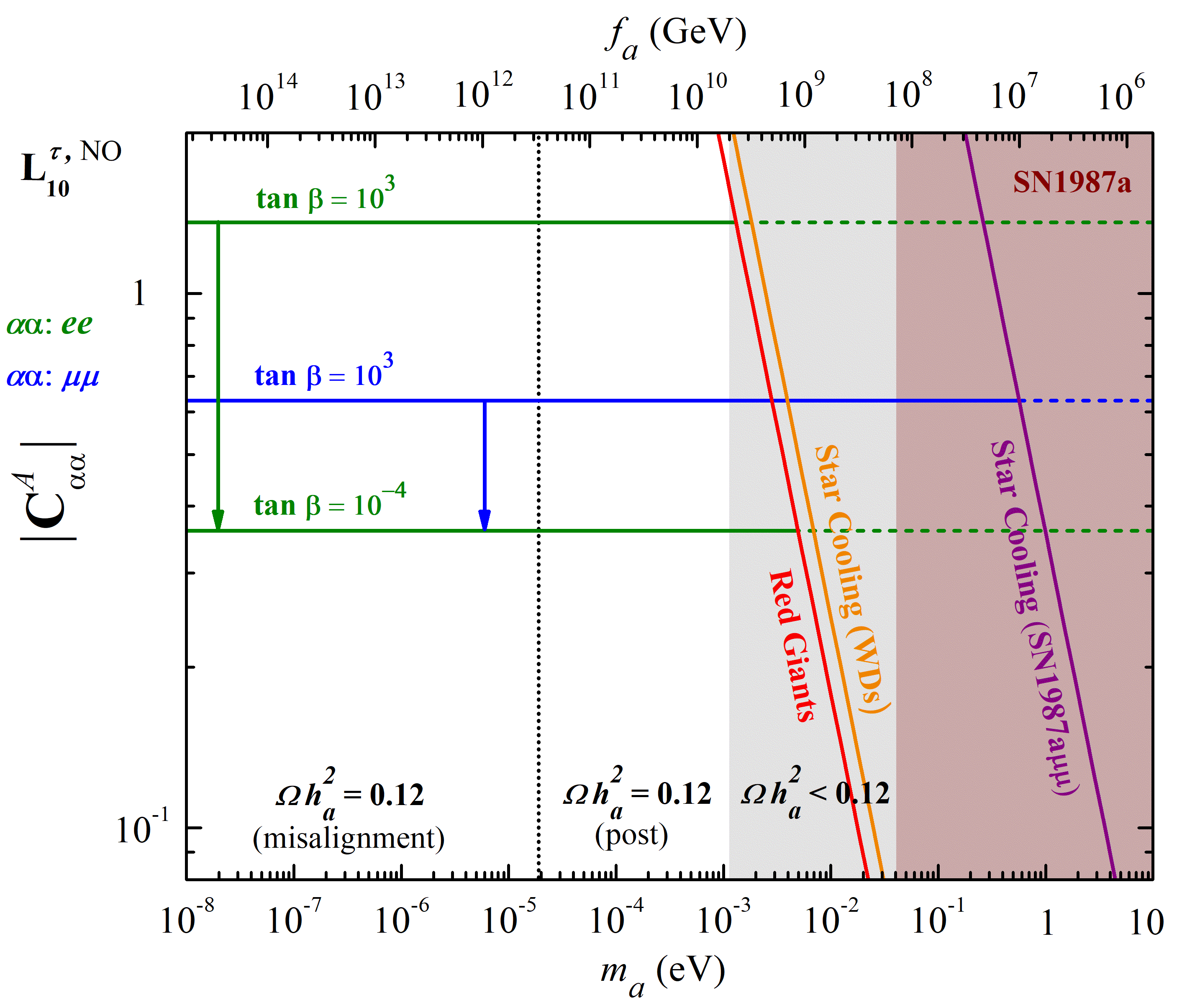}
    \caption{Axial diagonal axion-lepton couplings $|\mathbf{C}^A_{\alpha\alpha}|$, versus $m_a$ (bottom axis) and $f_a$ (top axis). We present the most restricted couplings $\alpha \alpha = ee$ ($\mu\mu$), indicated by horizontal green (blue) lines, for the lepton model $\text{L}_{10}^{\tau,\text{NO}}$ with $N_{\text{DW}} =1$ (see Table~\ref{tab:leptoncharges} and Sec.~\ref{sec:axiondarkmatter}). The dashed part of these lines are currently excluded by the red [purple] oblique bound from Red Giants~\cite{Capozzi:2020cbu,Bottaro:2023gep} [Star Cooling (SN1987$_a\mu\mu$)~\cite{MartinCamalich:2020dfe,Alonso-Alvarez:2023wig}]. We also indicate via an orange oblique line the constraint on the $ee$ coupling stemming from Star Cooling White Dwarfs~(WDs)~\cite{MartinCamalich:2020dfe,Alonso-Alvarez:2023wig} (see Table~\ref{tab:LeptonConstraints} and main text for details). The horizontal lines correspond to the $|\mathbf{C}^A_{\alpha\alpha}|$ values for the indicated $\tan \beta$. The remaining elements follow the color code of Fig.~\ref{fig:quarks}.}
    \label{fig:leptons}
\end{figure}
In the forthcoming numerical analysis, we consider the best fit values of the mass matrix elements for the models listed in Tables~\ref{tab:quarkcharges} and~\ref{tab:leptoncharges}, obtained through the $\chi^2$ analysis described in Sec.~\ref{sec:symmetries}. This enables the computation of the axion-fermion couplings through Eqs.~\eqref{eq:axionFermionCoupling}, which are then confronted with the most constraining present and future bounds, collected in Tables~\ref{tab:QuarkConstraints} and~\ref{tab:LeptonConstraints}. In Figs~\ref{fig:quarks} and~\ref{fig:leptons}, we present the most restricted couplings $\mathbf{C}^{V,A}_{\alpha \beta}$, indicated by horizontal lines, for some interesting quark and lepton minimal models, respectively, featuring $N_{\text{DW}} =1$. These results deserve several comments:
\begin{itemize}
    \item In models that are not $d$ or $s$ decoupled, $m_a$ is constrained by the $K^+ \rightarrow \pi^+ a$ decay as
    \begin{align}
        m_a \leq 
        2  \left(\frac{5.70\times 10^{12}\text{ GeV}}{\mathbf{F}^{V}_{ds}}\right)   |\mathbf{C}^{V}_{ds}  |^{-1}\;  \mu\text{eV} \;.
    \end{align}
    Considering the $\mathbf{F}^{V}_{ds}$ lower bounds in Table~\ref{tab:QuarkConstraints}, this relation exclude the regions on the right of the red oblique lines in Fig.~\ref{fig:quarks}. The values of the couplings $\mathbf{C}^V_{ds}$ for models $\text{Q}_{2}^{u,c,t}$ and $\text{Q}_{4}^{b}$ are shown by green horizontal lines. In these scenarios, the axion mass $m_a$ lies within the range $[10^{-5}, 10^{-4}]$ eV. Interestingly, for these cases the more predictive post-inflationary axion DM scenario (see Sec.~\ref{sec:axiondarkmatter}) is almost excluded (white region to the right of the vertical black-dotted line).

    Considering the axion mass bound, when combining the quark and lepton models permitting $N_{\text{DW}} = 1$, as shown in Table~\ref{tab:LeptonVSQuarkMassBound}, it becomes clear that the lepton constraints do not restrict the $\text{Q}_{2}^{u,c,t}$ and $\text{Q}_{4}^{b}$ models.
    
    \item For $d$ or $s$ decoupled models, the $K^+ \rightarrow \pi^+ a$ limit is automatically satisfied. As shown in Fig.~\ref{fig:quarks}, where the couplings $\mathbf{C}^V_{ds (uc)}$ for the models $\text{Q}_{4}^{d,s}$ are represented by  blue (orange) horizontal lines, other quark constraints impose a bound on $m_a$ comparable to that set by SN1987a, which excludes masses above $4 \times 10^{-2}$ eV~\cite{Carenza:2019pxu}.

    \item In Fig.~\ref{fig:leptons} we show the $\mathbf{C}^A_{ee,\mu\mu}$ limits for $\text{L}_{10}^\tau$ (the results for the remaining lepton models follow the same pattern). The most stringent bounds come from Red Giants being comparable to one stemming from Star Cooling (WDs), which constrains $\mathbf{C}^A_{ee}$ and imply  
    \begin{align}
        m_a \leq 
        2  \left(\frac{5.70\times 10^{12}\text{ GeV}}{\mathbf{F}_{ee}}\right) \left|\dfrac{s_\beta^2}{N} + \mathbf{C}^A_{ee}(\beta=0)  \right|^{-1} \; \mu\text{eV} \; ,
        \label{Eq:LeptonConservingContraint}
    \end{align}
    where we take the $\mathbf{F}_{ee}$ bounds from Table~\ref{tab:LeptonConstraints}. Note that this bound on $m_a$ depends on the angle $\beta$, as it arises from a flavor-conserving constraint. Therefore, we must account for the limits on $\beta$ imposed by perturbativity of Yukawa couplings. Requiring $|(\mathbf{Y}^f_{1,2})_{\alpha \beta}| \leq \sqrt{4\pi}$ leads to the following lower and upper limits on $t_\beta$:
    \begin{align}
        t^2_\beta \leq \frac{2\pi v^2}{|(\mathbf{M}_1^f)_{\alpha \beta}|^2} - 1,
        \quad
        t^2_\beta \geq {1}/{\left( \frac{2\pi v^2}{|(\mathbf{M}_2^f)_{\alpha \beta}|^2} - 1 \right)} \; ,
        \label{Eq:YukawaPerturbativity}
    \end{align}
    where, without loss of generality, we assume $\beta$ lies in the first quadrant. We have used the notation $\mathbf{M}_1^f$ ($\mathbf{M}_2^f$) for a mass matrix element originated from $\mathbf{Y}^f_{1}$ ($\mathbf{Y}^f_{2}$). At the end, the upper and lower bounds on $t_\beta$ are determined by the maximum values of $|(\mathbf{M}_1^f)|$ and $|(\mathbf{M}_2^f)|$, respectively. Using the above equation and the results of the $\chi^2$ fit for the mass matrix elements, we find that the quark Yukawas impose the most stringent bounds on $t_\beta$, with models $\text{Q}_{4}^{d,s}$ yielding $0.05 \leq t_\beta \leq 3.5$.
    Consequently, for a given set of $d,s$ decoupled mass matrices, the $m_a$ upper bound is within the range obtained from Eq.~\eqref{Eq:LeptonConservingContraint} considering the allowed interval for $\beta$. This contrasts with the cases $\text{Q}_{2}^{u,c,t}$ and $\text{Q}_{4}^{b}$ discussed above.

\end{itemize}

In Table~\ref{tab:LeptonVSQuarkMassBound} we show the $m_a$ bounds obtained when combining quark and lepton models with $N_{\text{DW}} =1$. We conclude that lepton constraints, rather than quark ones, set the axion mass bound in the $d$ or $s$ decoupled models within the range $m_a \in [10^{-3},10^{-2}]$ eV. Thererfore, the axion mass can be up to two orders of magnitude larger than in the previously analyzed scenarios. As a result, the whole post-inflationary region remains viable while still accommodating flavor-violating axion couplings. Although we have only covered flavor-conserving lepton constraints, we have checked that the lepton flavor-violating ones are too weak to significantly bound the axion mass.
\begin{table}[t!]
\renewcommand*{\arraystretch}{1.5}
\begin{tabular}{|c|c c c c c c c c c|}
\multicolumn{10}{c}{\textbf{Axion mass upper bound (meV)}}
\\
\hline
 \hline
  $\text{Q}_\text{4}^{d,s}$ & $[1.3,4.1]$ & $[0.6,0.9]$ & $[0.7,1.1]$ & $[1.3,3.9]$ & $[0.9,1.6]$ & $[0.7,1.1]$ & $[1.8,23.6]$ & $[1.7,11.4]$ & $[1.4,4.9]$ \\
 \hline
\multicolumn{1}{|c|}{} & \multicolumn{9}{c|}{For all lepton models above} \\
 \hline
$\text{Q}_\text{2}^u \; ; \;\text{Q}_\text{2}^c \;; \;\text{Q}_\text{2}^t \; ; \; \text{Q}_\text{4}^b$ & \multicolumn{9}{c|}{$3.7\times10^{-2} \;; \; 2.1\times10^{-2} \;; \; 1.5\times10^{-1} \;; \; 7.8\times10^{-2}$} \\
\hline
\end{tabular}
\caption{Upper bounds for $m_a$ (in meV) extracted from constraints on axion-to-fermion couplings of Tables~\ref{tab:QuarkConstraints} and~\ref{tab:LeptonConstraints}, for NO models of Tables~\ref{tab:quarkcharges} and~\ref{tab:leptoncharges} with $N_{\text{DW}}=1$. As explained in the main text, for $\text{Q}_\text{4}^{d,s}$ the bounds depend on $t_\beta$ which, taking into account Yukawa perturbativity [see Eq.~\eqref{Eq:YukawaPerturbativity}], reflect on the ranges given in the second row of the table for each lepton model.}
\label{tab:LeptonVSQuarkMassBound}
\end{table}
%

\section{Concluding remarks}
\label{sec:concl}

In this work we studied minimal $\nu$DFSZ axion models with flavored PQ symmetries and type-I seesaw neutrino masses. We performed a systematic analysis to identify the minimal models, in which the PQ symmetries impose the most restrictive quark and lepton flavor patterns, that are compatible with the observed fermion masses, mixing angles and CP-violating phases. We found eight maximally-restrictive quark flavor textures and eleven for the lepton sector for a combined eighty-eight possible models. In all cases, the number of independent parameters matches the number of observables. By construction, one neutrino is massless and, consequently, our framework can be tested at upcoming experiments looking for $(\beta \beta)_{0\nu}$ such as SNO+ II, LEGEND, and nEXO. We have also investigated several aspects related with axion phenomenology for each model, namely axion DM production in pre and post-inflationary cosmology. The most predictive scenarios are those which are free from DW problem. Namely, for $N_{\text{DW}}=1$ within the post-inflationary case, numerical simulations of axion strings restrict the axion decay constant to the interval $5\times 10^9$ GeV to $3 \times 10^{11}$ GeV. In Fig.~\ref{fig:gagg}, we show that  helioscopes and haloscopes are able to probe our models via their distinct axion-to-photon coupling predictions (see Table~\ref{tab:EN}). Future experiments such as IAXO, ADMX and MADMAX, will further scrutinize the models presented here. We also investigated how axion-to-fermion flavor violating couplings are constrained. We show that all our models, thanks to minimal flavored PQ symmetries, provide a natural framework to suppress flavor-violating couplings, as well as FCNCs in the Higgs sector. In fact, these symmetries lead to decoupled fermion states in the mass matrices, which impose restricted patterns in the $\mathbf{N}_f$ ($f=d,u,e$) matrices that control Higgs FCNCs. In particular, zero off-diagonal $\mathbf{N}_f$ entries, that otherwise contribute to the highly constraining flavor processes. This mechanism is also reflected in the vector and axial couplings of the axion to fermions (see Tables~\ref{tab:QuarkConstraints} and~\ref{tab:LeptonConstraints}). We find that in the quark sector the most stringent constraint is set by $K^+ \rightarrow \pi^+ + a$ in the $ds$ couplings, being automatically satisfied for models with a down or strange quark decoupled state. Strikingly, as shown in Fig.~\ref{fig:quarks}, for certain models with $N_{\text{DW}}=1$, the post-inflationary possibility for axion DM is excluded by this decay. In the lepton sector, current experimental flavor-violating constraints are not as relevant as those stemming from Red Giants and Star Cooling. These restrict the diagonal $ee$ and $\mu\mu$ axion-lepton couplings which depend on the VEV ratio of the two-Higgs doublets $\tan \beta$, providing a complementary constraint to that coming  from Higgs physics and scalar mediated FCNCs contributing to rare quark and lepton processes. For models with down or strange quark decoupled we found that the lepton Red Giants and Star Cooling constraints set a stronger bound on the axion mass (or scale) than the quark constraints -- see Table~\ref{tab:LeptonVSQuarkMassBound}.

To conclude, in this paper we have shown that flavored PQ symmetries in the minimal $\nu$DFSZ provide an appealing framework to address the flavor puzzle in the quark and lepton sector, neutrino mass generation, the strong CP problem and DM. This opens the possibility for further studies using axion frameworks as a gateway to tackle open problems in cosmology, particle and astroparticle physics.

\begin{acknowledgments}
We thank E.~Vitagliano for private communications. This research is supported by Fundação para a Ciência e a Tecnologia (FCT, Portugal) through the projects CFTP FCT Unit UIDB/00777/2020 and UIDP/00777/2020, CERN/FIS-PAR/0019/2021 and 2024.02004.CERN, which is partially funded through POCTI (FEDER), COMPETE, QREN and EU. The work of H.B.C. is supported by the PhD FCT grant 2021.06340.BD. 
\end{acknowledgments}



\end{document}